%% file: main.tex
\documentclass[final,11pt]{elsarticle}

\makeatletter
\def\ps@pprintTitle{%
    \let\@oddhead\@empty
    \let\@evenhead\@empty
    \let\@evenfoot\@oddfoot
    }
\makeatother

\usepackage{amssymb}
\usepackage{amsthm}

\usepackage{lineno}

\usepackage{amsmath}
\usepackage{amsbsy}
\usepackage{amsfonts}
\usepackage{bm}
\usepackage{bbm}
\usepackage[margin=1in]{geometry}
\usepackage{hyperref}
\usepackage{subfig}
\usepackage{dsfont}
\usepackage{mathtools}
\usepackage{multirow}

\usepackage{algorithm}
\usepackage{algpseudocode}

\usepackage{tikz}
\usepackage{standalone}

\theoremstyle{plain}

\theoremstyle{definition}

\theoremstyle{remark}

\usepackage{xcolor}
\definecolor{utorange}{RGB}{191,87,0}
\definecolor{utblue}{RGB}{0,169,183}

\usepackage[acronym]{glossaries}
\loadglsentries[\acronymtype]{acro.tex}

\journal{Journal of Computational Physics}

\begin{document}

\input{frontmatter}


\input{sections/introduction}
\input{sections/methods}
\input{sections/data_case_study}
\input{sections/ivygap_case_study}
\input{sections/conclusion}



\clearpage
\bibliographystyle{elsarticle-num}
\bibliography{references.bib}

\clearpage
\input{sections/appendix}






\end{document}

%% file: acro.tex
\newacronym{pde}{PDE}{partial differential equation}
\newacronym{mri}{MRI}{magnetic resonance imaging}
\newacronym{pto}{PtO}{parameter-to-observable}
\newacronym{mcmc}{MCMC}{Markov chain Monte Carlo}
\newacronym{map}{MAP}{maximum \textit{a posteriori}}
\newacronym{lq}{LQ}{linear-quadratic}
\newacronym[longplural=regions of interest]{roi}{ROI}{region of interest}
\newacronym{fem}{FEM}{finite element method}
\newacronym{adc}{ADC}{apparent diffusion coefficient}
\newacronym{dwi}{DWI}{diffusion-weighted imaging}
\newacronym{cgal}{\texttt{CGAL}}{Computational Geometry Algorithms Library}
\newacronym[longplural=quantities of interest]{qoi}{QoI}{quantity of interest}
\newacronym[longplural=degrees of freedom]{dof}{DOF}{degree of freedom}
\newacronym{la}{LA}{Laplace approximation}
\newacronym{cg}{CG}{conjugate gradient}
\newacronym{ccc}{CCC}{concordance correlation coefficient}
\newacronym{tacc}{TACC}{Texas Advanced Computing Center}

%% file: frontmatter.tex
\begin{frontmatter}



    \title{Predictive Digital Twins with Quantified Uncertainty for Patient-Specific Decision Making in Oncology}


    \author[Oden]{Graham Pash\corref{pash}}
    \ead{gtpash@utexas.edu}
    \cortext[pash]{Corresponding author.}
    
    \author[Oden]{Umberto Villa}
    \author[Oden,Livestrong]{David A. Hormuth II}
    \author[Oden,Livestrong,BME,DM,IP]{Thomas E. Yankeelov}
    \author[Oden]{Karen Willcox}
    
    \affiliation[Oden]{organization={Oden Institute for Computational Engineering and Sciences, The University of Texas at Austin},
        city={Austin},
        state={TX},
        postcode={78712},
        country={USA},
        }

    \affiliation[Livestrong]{organization={Livestrong Cancer Institutes, The University of Texas at Austin},
        city={Austin},
        state={TX},
        postcode={78712},
        country={USA},
        }
    
    \affiliation[BME]{organization={Department of Biomedical Engineering, The University of Texas at Austin},
        city={Austin},
        state={TX},
        postcode={78712},
        country={USA},
        }

    \affiliation[DM]{organization={Department of Diagnostic Medicine, The University of Texas at Austin Cancer Institutes},
        city={Austin},
        state={TX},
        postcode={78712},
        country={USA},
    }

    \affiliation[IP]{organization={Department of Imaging Physics, The University of Texas M.D. Anderson Cancer Center},
        city={Houston},
        state={TX},
        postcode={77230},
        country={USA},
        }
    
    \begin{abstract}
        Quantifying the uncertainty in predictive models is critical for establishing trust and enabling risk-informed decision making for personalized medicine. In contrast to one-size-fits-all approaches that seek to mitigate risk at the population level, digital twins enable personalized modeling thereby potentially improving individual patient outcomes. Realizing digital twins in biomedicine requires scalable and efficient methods to integrate patient data with mechanistic models of disease progression. This study develops an end-to-end data-to-decisions methodology that combines longitudinal non-invasive imaging data with mechanistic models to estimate and predict spatiotemporal tumor progression accounting for patient-specific anatomy. Through the solution of a statistical inverse problem, imaging data inform the spatially varying parameters of a reaction-diffusion model of tumor progression. An efficient parallel implementation of the forward model coupled with a scalable approximation of the Bayesian posterior distribution enables rigorous, but tractable, quantification of uncertainty due to the sparse, noisy measurements. The methodology is verified on a virtual patient with synthetic data to control for model inadequacy, noise level, and the frequency of data collection. The application to decision-making is illustrated by evaluating the importance of imaging frequency and formulating an optimal experimental design question. The clinical relevance is demonstrated through a model validation study on a cohort of patients with publicly available longitudinal imaging data.
    \end{abstract}




\end{frontmatter}

%% file: sections/introduction.tex
\section{Introduction}
\label{sec:intro}

\begin{figure*}[!htb]
    \centering
    \includegraphics[width=0.98\textwidth]{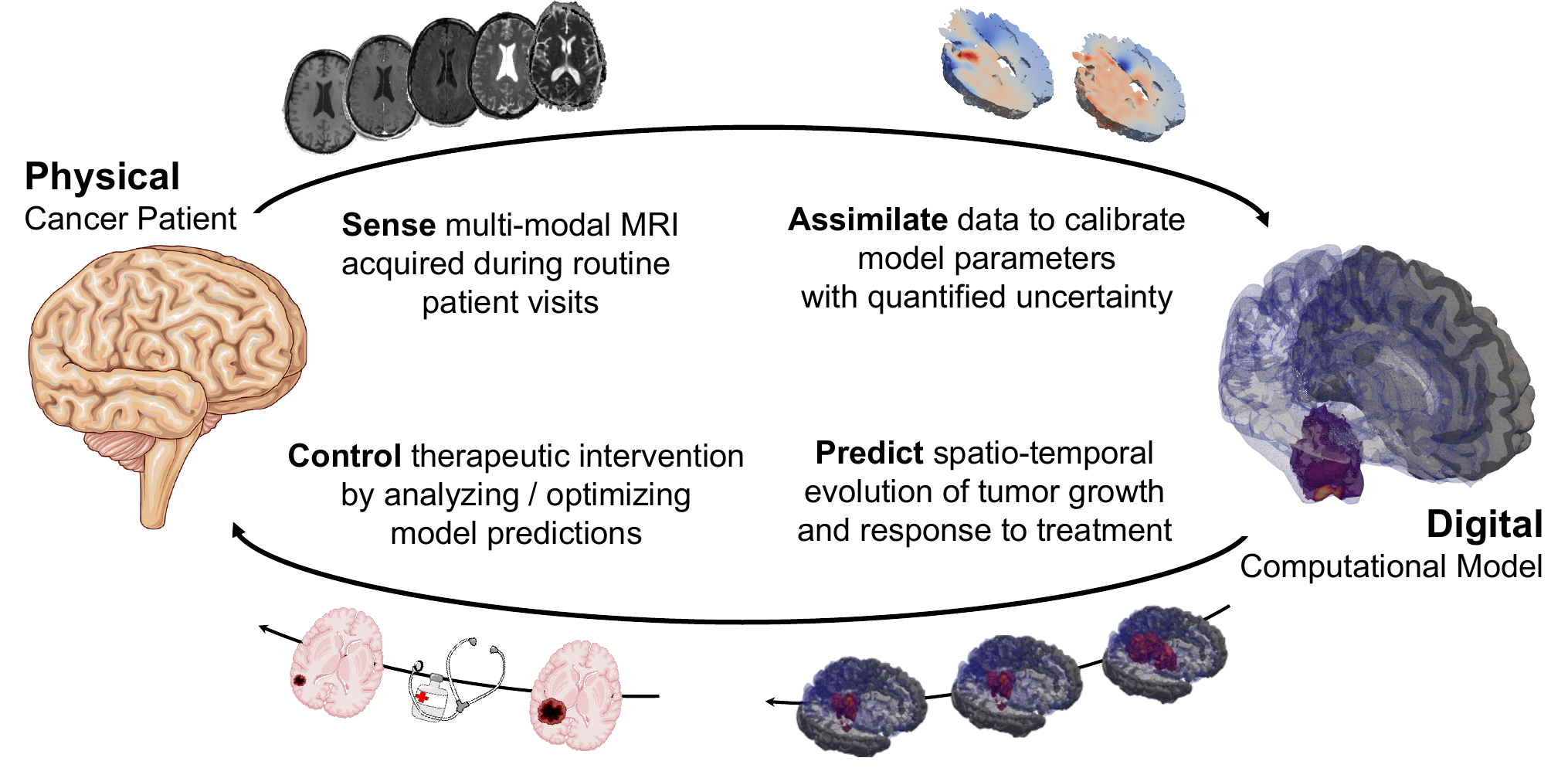}
    \caption{Illustration of our digital twin workflow for a cancer patient. Observational data are integrated with mechanistic models of tumor growth to update the computational representation of the dynamics with quantified uncertainty. The calibrated model is used to make probabilistic forecasts of tumor progression accounting for patient response to therapy. In turn, these forecasts guide clinical decision making. The model may be used to assess what-if scenarios for alternative interventions or to optimize therapy directly by (for example) adjusting dose level or schedule.}
    \label{fig:dtwin-overview}
\end{figure*}

Predictive digital twins are poised to make an impact in the burgeoning field of precision medicine by coupling mathematical models with patient-specific data. While it is not feasible to perform multiple \textit{in vivo} trials on an individual patient, mathematical and computational modeling augment the traditional clinical trial approach by enabling \textit{in silico} trialing and assessment of potential interventions in a personalized manner \cite{national2024foundational}. The heterogeneity in cancer physiology \cite{omuro2013glioblastoma} and patient response to a particular therapy \cite{aum2014molecular,hill2015hypoxia} means that there are many interventions that may work on average for a population, but are ineffective for an individual \cite{yankeelov2024designing}. Moreover, assessment of patient-specific response in the clinical setting is dependent on monitoring radiological and clinical changes over long timelines after the conclusion of therapy \cite{wen2010updated,chukwueke2019use}. An effectively realized digital twin will account for inter-patient variability in disease presentation and progression, while accelerating the feedback loop from data acquisition to clinical action \cite{hernandez2021digital,wu2022integrating,taylor2023patient,laubenbacher2024digital}. Figure~\ref{fig:dtwin-overview} illustrates the bidirectional flow of information between a cancer patient and their digital twin. Observational data are collected and combined with mechanistic models of the spatio-temporal tumor growth to update forecasts of prognosis, which are in turn used to guide clinical decisions between visits. In this work, we develop a scalable mathematical formulation and efficient computational framework of this digital twin data-to-decisions feedback loop, demonstrated by application to high-grade gliomas.

High-grade gliomas comprise the majority of malignant brain tumor cases and are characterized by their aggressive, invasive, and heterogeneous nature \cite{wang2013understanding}. Despite aggressive standard-of-care treatment \cite{stupp2005radiotherapy} consisting of multi-modal treatments (surgical resection, radiotherapy, and chemotherapy), overall prognosis remains poor and survival rates remain low \cite{tan2020management}. For a digital twin to be effective in this setting, the underlying models incorporating biological mechanisms must be able to reliably predict the complex spatio-temporal tumor dynamics. In particular, the variability of intra-tumoral properties presents a significant modeling challenge, requiring a spatial, and often high-dimensional, characterization of the model parameters driving growth to account for the variety of observed dynamics. One successful line of work employs image-based modeling to resolve the spatio-temporal tumor dynamics and leverages non-invasive medical imaging to calibrate models of tumor growth \cite{hogea2008image,jackson2015patient,oden2016toward,scheufele2020image,mang2020integrated,hormuth2021image,jarrett2021quantitative}. These data can be collected where patients receive care, both within and outside of academic research oriented facilities \cite{copur2016recap}, indicating the potential for broad impact. However, measurement noise, scarcity of observational data, and inadequacy of models to capture the biological complexity of tumor growth all contribute to significant modeling uncertainty \cite{hawkins2013bayesian}.

Uncertainty quantification plays a critical role in the development of a digital twin, helping establish trust in models and enabling risk estimation for robust decision making \cite{kouri2016risk}. The high-consequence nature of decisions in personalized medicine underscores the need for mathematical tools to rigorously account for uncertainty. There is a growing literature on the Bayesian calibration of image-based mechanistic models of tumor growth \cite{le2016mri,lima2017selection,lipkova2019personalized,chaudhuri2023predictive,liang2023bayesian}. However, the computational complexity of characterizing the posterior with expensive, nonlinear forward models remains a key challenge \cite{ghattas2021learning}. Low-dimensional representations of model parameters enable tractable exploration of the posterior and have shown promise in assessing alternative treatment plans in human patients \cite{lipkova2019personalized}. An alternative approach approximates the posterior directly, maintaining spatial resolution in the model parameters and demonstrating impressive predictive power in a murine model of brain cancer \cite{liang2023bayesian}. It remains to be shown that model parameterizations retaining sufficient spatial resolution to represent intra-tumoral heterogeneity can be scaled to the complex anatomies and long time horizons in human brain cancers. We address these modeling and computational challenges, with a particular focus on the advancement of scalable algorithms and efficient model implementations to enable the solution of high-dimensional Bayesian inverse problems in computational oncology.

Our central contribution is the development of a mathematical formulation for an end-to-end Bayesian data-to-decisions pipeline to serve as the basis for the development of digital twins in oncology. Clinical \gls*{mri} data are used to generate patient-specific computational geometries and extract information about the tumor biology. A computational framework is established for tractable Bayesian calibration of high-fidelity three-dimensional spatio-temporal models of tumor growth to the clinical data. Efficient, parallel, and flexible finite element representations of the governing equations are paired with state-of-the-art optimization algorithms. We numerically verify the ability to calibrate high-dimensional model parameters to \gls*{mri} data by reconstructing known spatial heterogeneity in a virtual patient. Furthermore, we investigate the impact of alternative imaging schedules on predictive performance and pose the optimal experimental design question of \textit{when} to observe. We demonstrate clinical relevance through a validation study on clinical data, identifying key aspects of model inadequacy.

The remainder of this paper is structured as follows. Section~\ref{sec:modeling} introduces the mathematical model of tumor growth and discusses the image processing pipeline. The scalable Bayesian methodology is presented in Section~\ref{sec:bip}. Section~\ref{sec:upenn} verifies the inverse problem methodology through an \textit{in silico} case study, illustrates the connection to clinical decision making, and formulates an optimal experimental design question. Section~\ref{sec:ivygap} establishes the real world utility of the framework applied to a cohort of patients with publicly available clinical data. The model's predictive ability is quantified along with potential limitations and extensions. Conclusions are drawn in Section~\ref{sec:conclusion}.

%% file: sections/methods.tex
\section{Image-based mechanistic modeling of tumors}
\label{sec:modeling}
This section provides necessary background on the mathematical modeling of tumor growth as well as the medical imaging data used for calibration. The governing \gls*{pde} model is derived in Section~\ref{sec:models} detailing tumor invasion, proliferation, and response to chemoradiation. Section~\ref{sec:data} describes a computational pipeline incorporating longitudinal image registration, tumor state estimation, and generation of a patient-specific computational geometry in the context of modeling glioma growth in the human brain.

\subsection{Mechanistic modeling of high-grade gliomas}
\label{sec:models}
Mathematical models of tumor growth primarily focus on two key characteristics: invasion of the tumor into the surrounding healthy tissue \cite{alfonso2017biology} and the proliferation of the existing tumor \cite{chaplain1996avascular}. Tumor invasion is typically modeled as a diffusion process and proliferation is modeled as a logistic growth process subject to a biological or physical carrying capacity. These phenomenological assumptions give rise to a semi-linear parabolic reaction-diffusion \gls*{pde} \cite{murray2002mathematical,murray2003mathematical}, which has proven successful in modeling the growth of solid tumors in a variety of organs \cite{clatz2005realistic,harpold2007evolution,garg2008preliminary,hormuth2015predicting}. Furthermore, it is common to assume a known carrying capacity of the tissue and model the tumor cellularity as a volume fraction. Henceforth, tumor cellularity and tumor volume fraction are used interchangeably unless explicitly noted. The model is given by:
\begin{equation}
\label{eqn:rd-model}
    \begin{alignedat}{2}
        \frac{\partial u}{\partial t} - \nabla\cdot\left(D\nabla u\right) - \kappa u(1-u) & = f \quad &  & \text{in $\Omega\times (t_0, t_f)$} \\
        u(x, t_0) & = u_0 & & \text{in $\Omega$} \\
        \nabla u \cdot \eta & = 0 & & \text{on $\partial\Omega\times (t_0, t_f)$}
    \end{alignedat}
\end{equation}
Here, $u(x, t)$ is tumor volume fraction which varies in time $t$ and over spatial coordinates $x$, $D(x)$ is the diffusion coefficient field, $\kappa(x)$ is the proliferation rate coefficient field, $(t_0,t_f)$ is the simulation window, $\Omega\subset \mathbb{R}^{n}$ is the bounded spatial domain with dimension $n \in \{2,3\}$, and $\eta$ is the outward unit normal to the boundary $\partial\Omega$. The homogeneous Neumann boundary condition stems from the assumption that the tumor does not grow beyond the boundary of the domain, for example, the skull. The initial cellularity $u_0(x)$ is determined from \gls*{mri} data following the procedure in Section~\ref{sec:data}. The source term $f$ is used to model the effect of treatment.

The standard-of-care therapy for high grade gliomas incorporates both highly conformal radiotherapy and systemic chemotherapy \cite{stupp2005radiotherapy}. Let $\mathcal{T}_{\text{rt}} = \{\tau_{k,\text{rt}}\}_k$ denote the collection of times at which radiotherapy is applied. Similarly, let $\mathcal{T}_{\text{ct}} = \{\tau_{k,\text{ct}}\}_k$ be the collection of times at which chemotherapy is administered. Together these define the treatment regimen. We make the common modeling assumption that the radiotherapy effect is instantaneous, killing some cells at the moment of treatment and with no lasting or time-delayed effects. This assumption results in a model of the form
\begin{equation}
    \label{eqn:radio}
    f_{\text{rt}}(u, z_{\text{rt}}, t) =
    \begin{cases}
        0 &\text{for $t\notin\mathcal{T}_{\text{rt}}$,}\\
        -\gamma\, (1 - S_{\text{rt}}(z_{\text{rt}}))\,u &\text{for $t\in\mathcal{T}_{\text{rt}}$,}
    \end{cases}
\end{equation}
where $z_{\text{rt}}$ is the applied radiation dose, $\gamma$ is a positive parameter to ensure appropriate dimensionality \cite{borasi2016modelling}, and $S_{\text{rt}}$ is the surviving fraction of the tumor after application of the therapy. We compute the surviving fraction with the well-established \gls*{lq} model relating applied dosage to radiotherapy induced cell death \cite{douglas1976effect,mcmahon2018linear}. The \gls*{lq} model calculates the surviving fraction as
\begin{equation}
    S_{\text{rt}}(z_{\text{rt}}) = \exp(-\alpha_{\text{rt}} z_{\text{rt}} - \beta_{\text{rt}} z_{\text{rt}}^2),
\end{equation}
where $\alpha_{\text{rt}} >0$ and $\beta_{\text{rt}} > 0$ are parameters describing the radiosensitivity of the tissue \cite{rockne2009mathematical,hormuth2020forecasting}. To capture the effect of chemotherapy, we assume a known drug efficacy $\alpha_{\text{ct}} > 0$ and clearance rate $\beta_{\text{ct}} > 0$ and employ a decaying exponential model of the form \cite{jarrett2020evaluating},
\begin{equation}
    \label{eqn:chemo}
    f_{\text{ct}}(u, z_{\text{ct}}, t) =
    \begin{cases}
        0 &\text{for $t < \tau_{0,\text{ct}}$,}\\
        -\alpha_{\text{ct}}(z_{\text{ct}}) \sum_{k} \exp(-\beta_{\text{ct}} (t-\tau_{k,\text{ct}}))\, u &\text{for $t \geq \tau_{0,\text{ct}}$.} 
    \end{cases}
\end{equation}
The drug efficacy $\alpha_{\text{ct}}$ may be modulated by the spatial distribution of the drug concentration $z_{\text{ct}}$ when data are available as in \cite{jarrett2020evaluating}, though we choose to model the term as a constant surviving fraction acting homogeneously in the domain as in \cite{hormuth2021image}. Finally, we combine the two effects as the source term in Eq.~\ref{eqn:rd-model},
\begin{equation}
    \label{eqn:treatment-model}
    f(u, z_{\text{rt}}, z_{\text{ct}}, t) = f_{\text{rt}}(u,z_{\text{rt}},t) + f_{\text{ct}}(u,z_{\text{ct}},t).
\end{equation}

We employ the \gls*{fem} \cite{hughes2003finite,oden2006finite} for our spatial discretization. Among other advantages, \gls*{fem} admits flexibility in the spatial discretization and is able to conform to the complex geometries arising in biomedical applications. The variational formulation for the model Eq.~\eqref{eqn:rd-model} is derived in \ref{sec:appendix-derivatives} and the implicit Euler method is used for the time discretization. We discuss a computational pipeline for meshing human brain anatomy in the following section.

\subsection{Computational pipeline: From medical imaging to computational models}
\label{sec:data}
\begin{figure*}[!htb]
    \centering
    \includegraphics[width=0.95\textwidth]{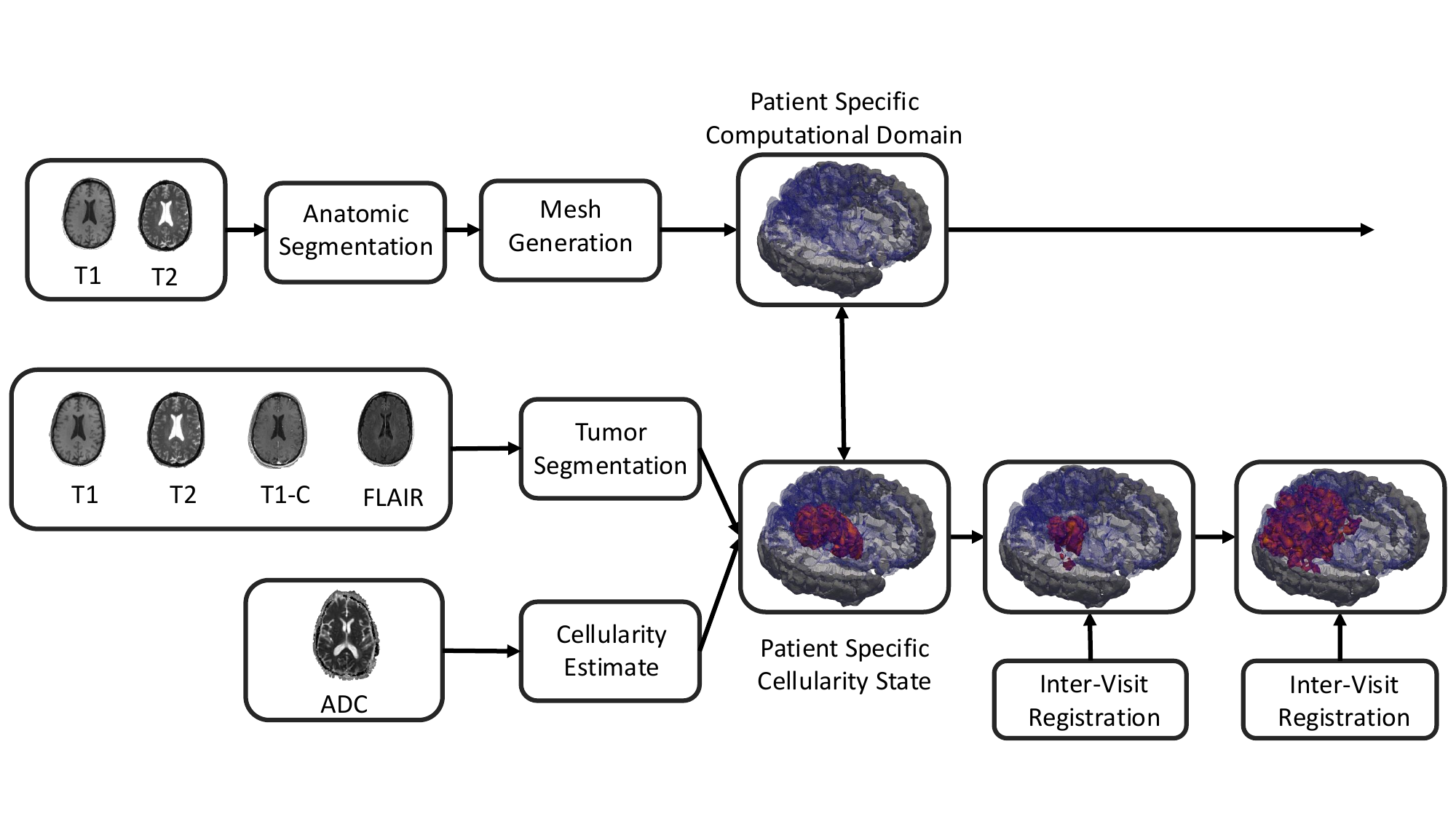}
    \caption{The computational pipeline: anatomic segmentation and mesh generation, cellularity estimation, and longitudinal registration. Together $T_1$- and $T_2$-weighted scans are used to generate a computational domain tailored to the patient's anatomy. Tumor cellular density is estimated by combining ADC imaging with tumor segmentations. When radiologist defined segmentations are not available, automated tools utilizing $T_1$ pre- and post-contrast imaging along with $T_2$-weighted and FLAIR modalities are used to develop appropriate regions of interest.}
    \label{fig:pipeline}
\end{figure*}

Biomedical imaging, especially \gls*{mri} \cite{yankeelov2007dynamic,padhani2009diffusion}, plays a critical role in the diagnosis, treatment planning, and management of a variety of tumors, such as those in the brain \cite{bernstock2022standard}, the breast \cite{karellas2008breast}, and the prostate \cite{hricak2007imaging}. In our digital twin framework, we consider this to be the primary source of observational data coming from the clinic. To integrate imaging data with the biophysical models described in Section~\ref{sec:models}, we must estimate the tumor volume fraction, i.e. the state variable of the \gls*{pde}, and define an appropriate computational domain. Figure~\ref{fig:pipeline} provides an overview of such a computational pipeline for the human brain.

Expert segmentation of tumor \glspl*{roi} are used to determine tumor extent, when available. However, semi-automated approaches leveraging routinely collected \gls*{mri} data are improving \cite{kofler2020brats} and may be used when expert segmentation is unavailable. The \gls*{adc} quantifies the rate of diffusion of water molecules, which is inversely proportional to cellular density. The \gls*{adc} can be used to estimate the tumor cellularity \cite{jarrett2021quantitative,ellingson2011spatially} via
\begin{equation}
\label{eqn:adc}
    d(\bar{x}, t) =  \frac{\text{ADC}_w - \text{ADC}(\bar{x},t)}{\text{ADC}_w - \text{ADC}_{\text{min}}},
\end{equation}
where ADC$_w$ is the apparent diffusion coefficient of free water \cite{hormuth2021image}, ADC$(\bar{x},t)$ is the measured apparent diffusion coefficient at time $t$ in the voxel with spatial coordinate $\bar{x}$, and ADC$_\text{min}$ is the minimum value recorded within the tumor. This approach generates observations of the tumor cellularity, $d(\bar{x}, t)$, every time the patient is imaged. To ensure a consistent frame of reference, longitudinally collected data is rigidly registered to a baseline \gls*{mri} using a tool such as \texttt{elastix} \cite{klein2009elastix,shamonin2014fast}. Similarly, multiple modalities within a visit may be registered using a mutual information criterion to account for varying image resolution and contrast \cite{klein2009elastix,thevenaz2000optimization}. For an overview of medical image registration, readers are referred to \cite{viergever2016survey}.

To showcase the approach in application to brain cancer, we consider two publicly available datasets: UPENN-GBM \cite{bakas2022university,bakas2021tcia} and IvyGAP \cite{puchalski2018anatomic,shah2016tcia}. Both datasets contain pre- and post-contrast $T_1$-weighted, $T_2$-weighted, and FLAIR images. The UPENN-GBM dataset also provides expert segmentation of the tumor \gls*{roi}, while a semi-automated approach \cite{juan2015automated} is used to generate tumor \glspl*{roi} for the IvyGAP dataset. Additionally, the IvyGAP dataset contains \gls*{dwi} and \gls*{adc} data as well as longitudinal imaging data, enabling retrospective studies with clinical data. Brief descriptions of collected modalities and their connection to biophysical modeling are summarized in Table~\ref{tab:mri}, adapted from \cite{mabray2015modern}.

\begin{table}[!htb]
    \centering
    \caption{Summary of acquired \gls*{mri} and their utility.}
    \begin{tabular}{l|l}
        Imaging technique & Major utility in brain tumor imaging \\
        \hline
        Pre- and post-contrast $T_1$ & Anatomy, necrosis, enhancing tumor \\
        $T_2$ / FLAIR & Anatomy, edema, non-enhancing tumor \\
        DWI / ADC & Cellularity estimation \\
    \end{tabular}
    \label{tab:mri}
\end{table}

To construct a reference computational geometry for the biological domain upon which to simulate the growth of the tumor, we build upon the work of \cite{mardal2022mathematical}. We use $T_1$- and $T_2$-weighted images for segmentation of the gray matter, white matter and cerebral aqueduct using the \texttt{FreeSurfer} neuroimaging software \cite{fischl2012freesurfer}. To ensure reliable segmentation when large lesions are present, the $T_1$ and $T_2$ images are first repaired with virtual brain grafting \cite{radwan2021virtual}. Tetrahedral meshing of the resultant surfaces is performed with \texttt{CGAL} \cite{fabri2009cgal}. Similar approaches have provided a basis for high-fidelity simulation of a variety of physical phenomena in the brain from the circulation of cebrospinal fluid \cite{croci2021fast,corti2023numerical,dreyer2024modeling} to neurodegenerative disease progression \cite{corti2023discontinuous,thompson2024alzheimer}.

\section{Scalable Bayesian model calibration}
\label{sec:bip}

A fundamental capability of a digital twin is the ability to integrate and assimilate observational data with mathematical and computational models. This section details a scalable methodology to estimate model parameters in the presence of uncertainty. We formulate this process as a Bayesian inverse problem, seek an approximation to the posterior distribution, and detail how to propagate uncertainty through the forward model to generate probabilistic forecasts of tumor growth and clinically relevant quantities of interest.

\subsection{Formulation}
\label{sec:bip_formulation}
We pose the calibration of the mechanistic model Eq.~\eqref{eqn:rd-model} to the imaging data as an inverse problem: from observational data $\boldsymbol{d} = [d(\bar{x}, t_1), d(\bar{x}, t_2), \dots, d(\bar{x}, t_{n_t})] \in\mathbb{R}^{n_d \times n_t}$, we infer the values of the unknown parameter(s) $m\in\mathcal{M}$ for some suitable function space $\mathcal{M}$. Here, $n_d$ is the dimensionality of a single data point, i.e., the number of voxels in an \gls*{mri} image, and $n_t$ is the number of visits at which tumor volume fraction estimates are acquired. We choose the parameterization $m = [m_D, m_\kappa] = [\log(D), \log(\kappa)]$ to preserve positivity of the model coefficients $D=\exp(m_D),\kappa=\exp(m_\kappa)$. The objective is to estimate the posterior distribution of the model parameters, given by Bayes' theorem \cite{stuart2010inverse} as:
\begin{equation}
    \label{eqn:bayes}
    \frac{d \nu_{\text{post}}}{d\nu_{\text{pr}}}\propto \pi_{\text{like}}(\boldsymbol{d} \vert m),
\end{equation}
where $d\nu_{\text{post}}/d\nu_{\text{pr}}$ denotes the Radon-Nikodym derivative \cite{williams1991probability} of the posterior measure $\nu_{\text{post}}$ with respect to the prior measure $\nu_{\text{pr}}$. The likelihood $\pi_{\text{like}}$ is specified by the choice of noise model.

\subsection{Likelihood}
\label{sec:likelihood}
To account for measurement uncertainty in the pipeline outlined in Section~\ref{sec:data}, we assume an additive Gaussian noise model,
\begin{equation}
    \label{eqn:noise-model}
    d(\bar{x}, t_i) = \mathcal{B} (u(x,t_i)) + \varepsilon_i,
\end{equation}
where $\mathcal{B}:\mathcal{U}\rightarrow\mathbb{R}^{n_d}$ is an observation operator that extracts the observable from the state $u\in\mathcal{U}$, for some suitable Hilbert space $\mathcal{U}$. For example, an observable may be the voxel-resolved tumor cellularity. Additionally, the observation operator appropriately interpolates between the spatial coordinates of the \gls*{fem} representation $x$ and the spatial coordinates of the voxel data $\bar{x}$. The additive Gaussian noise is assumed to have zero mean and covariance $\boldsymbol{\Gamma}_{\text{noise}}\in\mathbb{R}^{n_d \times n_d}$, such that $\varepsilon_i \sim\mathcal{N}(0,\boldsymbol{\Gamma}_{\text{noise}})$. Further, we assume uniformity of the noise, $\boldsymbol{\Gamma}_{\text{noise}} = \sigma_{\text{noise}}^2\mathbf{I}$, for some unknown noise variance $\sigma^2_{\text{noise}}$. We define the \gls*{pto} map:
\begin{equation}
\label{eqn:p2o}
    \mathcal{F}:\mathcal{M}\rightarrow\mathbb{R}^{n_d\times n_t} = [\mathcal{B}(u(x, t_1)), \mathcal{B}(u(x,t_2)), \dots, \mathcal{B}(u(x,t_{n_t}))]\,,\quad     \text{s.t.}\,\, r(u,m)=0,
\end{equation}
where $r(u,m)=0$ is the residual form of the governing \gls*{pde} Eq.~\eqref{eqn:rd-model}, explicitly noting the dependence on the state and parameter. Evaluating the \gls*{pto} map $\mathcal{F}$ involves the solution of the forward model Eq.~\eqref{eqn:rd-model} for the state $u\in\mathcal{U}$ and the application of the observation operator $\mathcal{B}$ to extract the observable at each time that data is acquired. The probability density function of the likelihood may be expressed as
\begin{equation}
    \label{eqn:likelihood}
    \pi_{\text{like}}(\boldsymbol{d} \vert m)\propto \exp\left\{-\Phi(m; \boldsymbol{d})\right\},
\end{equation}
where $\Phi(m; \boldsymbol{d})$ is the negative log-likelihood. For the additive Gaussian case with multiple measurements, we sum the negative log-likelihoods for each observation,
\begin{equation}
\label{eqn:nll}
    \Phi(m, \boldsymbol{d}) := \sum\limits_{i=1}^{n_t} \frac{1}{2}\|\mathcal{F}_i(m) - \boldsymbol{d}_i\|^2_{\boldsymbol{\Gamma}^{-1}_{\text{noise}}},
\end{equation}
where $\mathcal{F}_i$ denotes the $i$-th component of the \gls*{pto} map, that is, $\mathcal{F}_i := \mathcal{B}(u(x,t_i))$.

\subsection{Prior distribution}
\label{sec:prior}
The prior distribution encodes structural knowledge of the parameters and is taken to be a Gaussian random field, with mean $m_{\text{pr}}$ and covariance operator $\mathcal{C}_{\text{pr}}$, such that $m\sim\mathcal{N}(m_{\text{pr}},\mathcal{C}_{\text{pr}})$. This modeling choice has desirable properties for analysis and computation, and has been used to model various physical systems including tumor growth in murine models \cite{liang2023bayesian}. In particular, this selection ensures a well-posed posterior \cite{stuart2010inverse} and implies a prior measure of the form
\begin{equation}
    \label{eqn:prior}
    d\nu_{\text{pr}}(m) \propto \exp\left\{-\frac{1}{2}\|m - m_{\text{pr}}\|^2_{\mathcal{C}^{-1}_{\text{pr}}}\right\}.
\end{equation}
Values for the mean $m_{\text{pr}}$ may be taken from the literature or determined by deterministic calibration to a cohort.

The seminal work \cite{lindgren2011explicit} established a link between Mat\'ern covariance operators and precision operators defined by \glspl*{pde}. Following \cite{stuart2010inverse,villa2021hippylib} the covariance operator is taken to be the inverse of an elliptic operator of the form 
\begin{equation}
\label{eqn:covprior}
    \mathcal{C}_{\text{pr}}:=\mathcal{A}^{-2}=(-\gamma\Delta + \delta I)^{-2}
\end{equation}
with hyperparameters $\gamma>0$ and $\delta>0$ that control the correlation length $\rho$ and the pointwise variance $\sigma^2$ of the operator \cite{lindgren2011explicit}. Specifically, the hyperparameters are related by
\begin{equation}
\label{eqn:hyperparameters}
    \delta = \frac{\sqrt{2}}{\sigma\rho\sqrt{\pi}} \quad,\quad \gamma=\frac{\rho}{4\sigma\sqrt{2\pi}}.
\end{equation}
A Robin boundary condition is applied to the operator $\mathcal{C}_{\text{pr}}$ to reduce boundary artifacts \cite{daon2016mitigating}.

We assume the log coefficient field parameters $m$ to be independent and model each as a Gaussian random field as described above. That is, $m_D\sim\mathcal{N}(m_{\text{pr},D}, \mathcal{C}_{\text{pr},D})$ and $m_\kappa\sim\mathcal{N}(m_{\text{pr},\kappa}, \mathcal{C}_{\text{pr},\kappa})$. Thus, $m_\text{pr} := \left[m_{\text{pr},D}, m_{\text{pr},\kappa}\right]$ and $\mathcal{C}_{\text{pr}}$ has a block diagonal structure with blocks $\mathcal{C}_{\text{pr},D}$ and $\mathcal{C}_{\text{pr},\kappa}$, respectively.

\subsection{Low-rank Laplace approximation to the posterior distribution}
\label{sec:la-post}
We reiterate that we model the parameters as spatial fields which are high-dimensional upon discretization. For example, when using first-order Lagrange elements, each discretized parameter has size equal to the number of vertices in the mesh. The computational cost to thoroughly explore the posterior distribution with standard methods, such as \gls*{mcmc}, is prohibitive in this setting. We instead leverage the Laplace approximation to the posterior \cite{bui2013computational,isaac2015scalable} for a tractable solution,
\begin{equation}
\label{eqn:la}
    \nu_{\text{post}}^{\text{LA}} \propto \mathcal{N}(m_{\text{MAP}}, \mathcal{C}_{\text{post}}).
\end{equation}

The Laplace approximation is a Gaussian approximation to the posterior centered at the \gls*{map} point $m_{\text{MAP}}$ with covariance $\mathcal{C}_{\text{post}}$ equal to the inverse of the negative log-posterior Hessian evaluated at $m_{\text{MAP}}$. This approximation is exact in the case of linear inverse problems and will serve as a cost-effective surrogate for the nonlinear problem considered in this work. Building the Laplace approximation requires one to compute the \gls*{map} point and calculate the covariance.

We compute the \gls*{map} estimate by minimizing the negative log-posterior
\begin{equation}
    \label{eqn:map}
    m_{\text{MAP}} := \underset{m\in\mathcal{M}}{\operatorname{arg\,min}} \left(- \log\nu_{\text{post}}(m\vert \boldsymbol{d}) \right).
\end{equation}
Recalling the expression for the likelihood Eq.~\eqref{eqn:likelihood} and the prior Eq.~\eqref{eqn:prior}, we may write the posterior as
\begin{equation}
\label{eqn:posterior}
    \nu_{\text{post}}(m\vert \boldsymbol{d})\propto \left\{ \underbrace{-\sum_{i} \frac{1}{2}\|\mathcal{F}_i(m) - \boldsymbol{d}_i\|^2_{\boldsymbol{\Gamma}^{-1}_{\text{noise}}}}_{\text{data misfit}} - \underbrace{\vphantom{\sum_j} \frac{1}{2}\|m - m_{\text{pr}}\|^2_{\mathcal{C}^{-1}_{\text{pr}}}}_{\text{prior}} \right\}.
\end{equation}
Thus, computing the \gls*{map} point amounts to solving a deterministic inverse problem with the log-likelihood Eq.~\eqref{eqn:nll} playing the role of the data misfit term and the prior Eq.~\eqref{eqn:prior} acting as regularization.

Inexact Newton-Krylov algorithms have been shown to efficiently solve the optimization problem Eq.~\eqref{eqn:map} \cite{petra2012inexact}. Early termination of \gls*{cg} iterations is used to inexactly solve the Newton system using the Eisenstat-Walker criterion to prevent over-solving \cite{eisenstat1996choosing} and the Steihaug criterion to avoid negative curvature \cite{steihaug1983local}. Globalization is performed with an Armijo backtracking line search to guarantee global convergence \cite{nocedal1999numerical}. For a wide class of nonlinear inverse problems, the number of outer Newton iterations and inner \gls*{cg} iterations is independent of the mesh size and hence parameter dimension \cite{heinkenschloss1993mesh}. This is a consequence of the Newton solver, the compactness of the Hessian of the data misfit term in Eq.~\eqref{eqn:posterior}, and preconditioning by the inverse of the regularization operator. The gradient and Hessian action are obtained using the adjoint method, thereby limiting the number of expensive \gls*{pde} solves that are required. These expressions are derived in \ref{sec:appendix-derivatives} with the Lagrangian formalism \cite{troltzsch2010optimal,manzoni2021optimal}.

The covariance of the Laplace approximation to the posterior is given by the inverse of the Hessian of the negative log-posterior,
\begin{equation}
\label{eqn:covpost}
    \mathcal{C}_{\text{post}} := \mathcal{H}(m_{\text{MAP}})^{-1} = \left( \mathcal{H}_{\text{misfit}}(m_{\text{MAP}}) + \mathcal{C}_{\text{pr}}^{-1} \right)^{-1},
\end{equation}
where $\mathcal{H}_{\text{misfit}}$ denotes the Hessian of the negative log-likelihood. Upon discretization, we have
\begin{equation}
    \label{eqn:discrete-covpost}
    \mathbf{\Gamma}_{\text{post}} = \left( \mathbf{H}_{\text{misfit}}(m_{\text{MAP}}) + \mathbf{\Gamma}^{-1}_{\text{pr}} \right)^{-1}
\end{equation}
where $\mathbf{\Gamma}_{\text{post}}$ is the discretization of $\mathcal{C}_{\text{post}}$, $\mathbf{\Gamma}_{\text{pr}}$ is the discretization of $\mathcal{C}_{\text{pr}}$, and $\mathbf{H}_{\text{misfit}}(m_{\text{MAP}})$ is the discretized Hessian of the negative log-likelihood at the \gls*{map} estimate. The construction of $\mathbf{H}_{\text{misfit}}$ for large-scale applications is prohibitive, let alone the inversion required to form $\mathbf{\Gamma}_{\text{post}}$. Instead, we leverage the fact that in many cases, the spectral decay of $\mathbf{H}_{\text{misfit}}$ is rapid as the data contain limited information about the (infinite-dimensional) parameters. Thus we adopt a low-rank correction to the prior covariance to approximate the posterior as in \cite{isaac2015scalable}. Randomized algorithms \cite{halko2011finding} are used to estimate the leading eigenpairs $\{(\lambda_j, v_j)\}_i^k$ of the generalized eigenvalue problem
\begin{equation}
\label{eqn:GEVP}
    \mathbf{H}_\text{misfit}v_j = \lambda_j \boldsymbol{\Gamma}^{-1}_{\text{pr}}v_j.
\end{equation}
The eigenvectors associated with the dominant eigenvalues represent the directions in parameter space most informed by the data. The covariance matrix in Eq.~\eqref{eqn:covpost} is approximated using the Sherman-Morrison-Woodbury identity as in \cite{isaac2015scalable},
\begin{equation}
\label{eqn:lr-covpost}
    \boldsymbol{\Gamma}_{\text{post}} \approx \boldsymbol{\Gamma}_{\text{pr}} - \sum^k_{j=1} \frac{\lambda_j}{1+\lambda_j} v_jv_j^T.
\end{equation}
Since $k$ is much smaller and independent of the discretized dimension of the parameters, this method is a scalable approach for high-dimensional Bayesian inversion. We employ \texttt{hIPPYlib} \cite{villa2018hippylib,villa2021hippylib,kim2023hippylib} for implementation of the Newton-\gls*{cg} solver as well as the Mat\'ern prior.

\subsection{Propagation of uncertainty}
\label{sec:qoi}
Ultimately, we wish to use the model \eqref{eqn:rd-model} to \emph{predict} the future evolution of the tumor and compute meaningful \glspl*{qoi}. Furthermore, we are interested in the propagating uncertainty in the parameters through the forward model to generate predictive distributions of the \glspl*{qoi}. To this end, we define the prediction map, $\mathcal{F}_{\text{pred}}$, where the tumor state is computed at some final time $t_{\text{pred}}$,
\begin{equation}
\label{eqn:p2o-pred}
    \mathcal{F}_{\text{pred}}:\mathcal{M}\rightarrow\mathcal{U} := u(x,t_{\text{pred}})\,,\quad \text{s.t. } r(u,m; u_0, (t_0,t_{\text{pred}})) = 0.
\end{equation}
Once more $r$ denotes the \gls*{pde} model in residual form, this time explicitly noting the dependence on parameter $m$, simulation window $(t_0,t_{\text{pred}})$ and initial condition $u_0$. For example, in the prediction setting one may wish to estimate the initial condition $u_0$ from the last acquired \gls*{mri} observation and predict over a new simulation window representing, say, one month into the future. Further, consider a generic quantity of interest $q:\mathcal{U}\rightarrow\mathcal{Q}$ for some suitable space $\mathcal{Q}$. We are interested in computing the pushforward of a measure through the composition of the \gls*{qoi} map with the prediction map, $(q\circ\mathcal{F}_{\text{pred}})_\sharp \nu$. Here $\cdot_\sharp$ denotes the pushforward of a measurable function \cite{bogachev2007measure}. Mechanically, we sample $m\sim\nu$ and compute $q(\mathcal{F}_{\text{pred}}(m))$. This is called the prior predictive distribution when $\nu = \nu_{\text{pr}}$ and the posterior predictive distribution when $\nu = \nu_{\text{post}}$.

We consider four \glspl*{qoi}: total tumor cellularity, tumor volume, the \gls*{ccc}, and the Dice similarity coefficient. We first define a threshold for the measurable tumor region, $\hat{u}(x, t) := u(x, t) > \bar{u}$. The threshold $\bar{u}$ is the volume fraction at which the lesion is measurable and is taken to be $0.1$ for this study. The total tumor cellularity is computed as the integration of the measurable tumor volume fraction over the domain, $q_{\text{TTC}}:=\int_\Omega \hat{u}\,dx$, and may optionally be scaled by the carrying capacity to recover the number of tumor cells. The tumor volume is computed by integrating an indicator function for regions with measurable tumor over the domain, $q_{\text{TV}} := \int_\Omega \mathbbm{1}_{\hat{u}}\,dx$. The \gls*{ccc} \cite{lawrence1989concordance} measures correlation between two datasets, with a penalty for deviating from the line of unity. This is used to quantify voxel-wise agreement, for example, between the observed data and the model prediction. The Dice similarity coefficient \cite{dice1945measures,sorensen1948method} measures the degree of spatial overlap and is used to compare predicted measurable tumor with the observed tumor. The Dice coefficient is defined as $2\vert X \cap Y\vert / (\vert X \vert + \vert Y \vert)$ for two sets $X, Y$ with $\vert\cdot\vert$ denoting the cardinality, in this case the sets are the indicator functions for the predicted measurable tumor $\hat{u}$ and the true, observed tumor $u^\dagger$.

%% file: sections/data_case_study.tex
\section{Data-to-decisions: Demonstration on a virtual patient}
\label{sec:upenn}

As an initial demonstration of the digital twin bidirectional feedback loop illustrated in Figure~\ref{fig:dtwin-overview}, we deploy the methodology on a virtual patient. This setting allows for control of the experimental setup, including the underlying tumor growth mechanism, noise level, and frequency of observation. We first detail the data generation process and verify the inverse problem methodology in the controlled setting. Further, we assess the potential benefit of alternative imaging regimens on forecasts of tumor growth. To support the envisioned clinical deployment of the approach, we also establish the computational tractability.

\subsection{Synthetic data generation}
\begin{figure}[!htb]
    \centering
    \includegraphics[width=\linewidth]{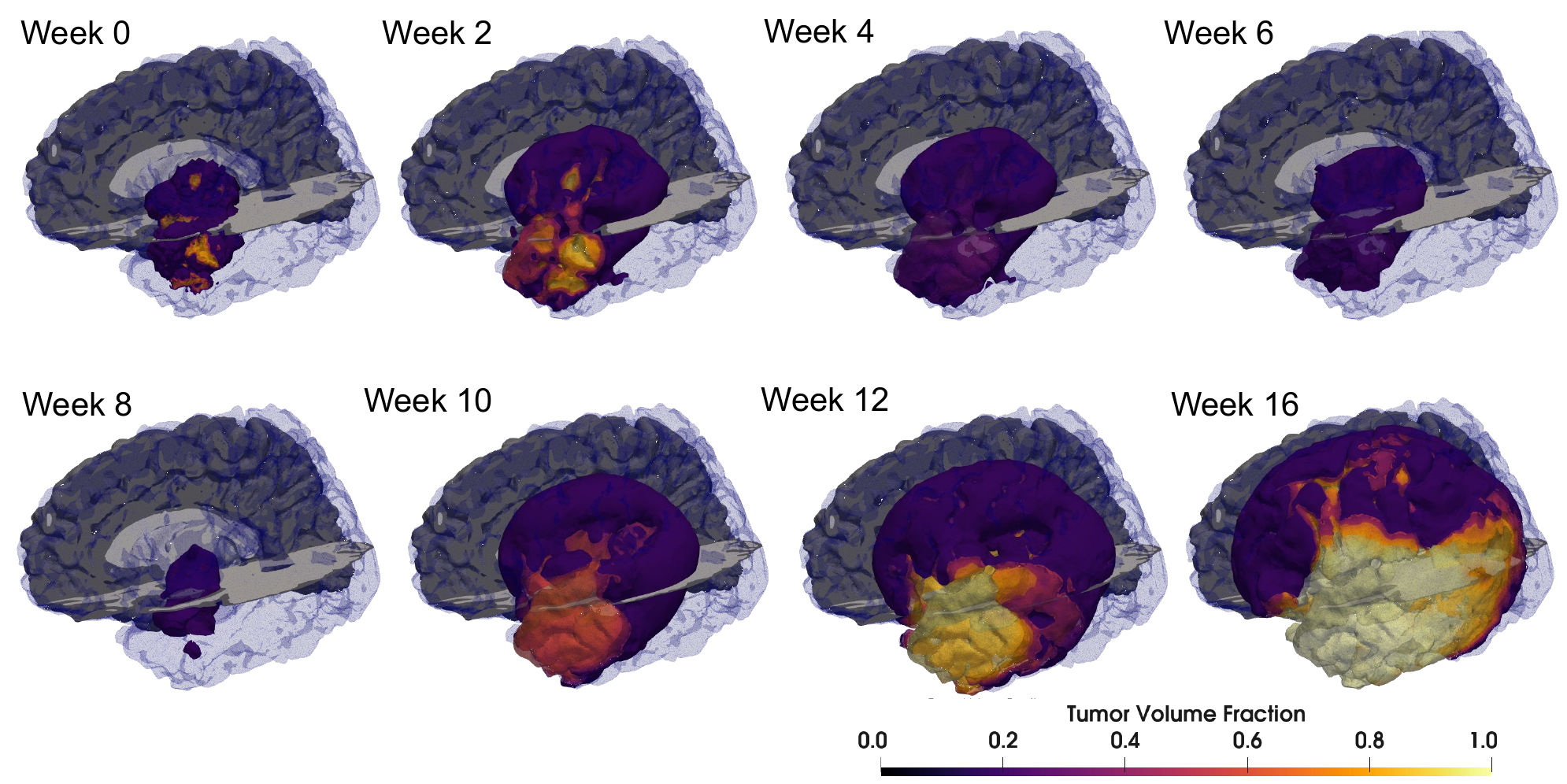}
    \caption{Snapshots of synthetic tumor progression for UPENN-GBM subject 101. Note the heterogeneous initial state and the retreat while under therapy (weeks 2-8). When therapy ends, tumor recurrence is swift and the tumor extent is larger at the prediction time (week 16) than during the imaging window (up to week 12).}
    \label{fig:sub-00101-snapshots}
\end{figure}

We utilize the publicly available UPENN-GBM dataset \cite{bakas2022university,bakas2021tcia}. In particular, we generate a computational geometry of subject \texttt{101} following the procedure outlined in Section~\ref{sec:data}. Observational data is synthesized by solving Eq.~\eqref{eqn:rd-model} forward from the initial condition for two weeks of untreated growth followed by the Stupp protocol \cite{stupp2005radiotherapy} (six weeks of concurrent radiotherapy and chemotherapy), followed by another month of untreated growth, that is, without adjuvant temozolomide. Three different scenarios are considered in which observational data are collected daily, weekly, or fortnightly. The prediction window is one month after the conclusion of therapy. Snapshots of the tumor progression are shown in Figure~\ref{fig:sub-00101-snapshots}.

The mesh used for data generation is comprised of $447,490$ vertices. First order Lagrange elements are used for the state, resulting in an equal number of \glspl*{dof}. Since the UPENN-GBM dataset lacks \gls*{adc} estimates, we follow \cite{swanson2008mathematical} and take the tumor volume fraction is taken to be $0.8$ and $0.16$ in the enhancing and non-enhancing regions of the tumor, respectively. The non-enhancing region is characterized by invasive and diffuse disease and the lower cellularity is chosen to account for the well-documented difficulties in separating proliferative tumor from other processes, such as vasogenic edema \cite{puchalski2018anatomic}. The underlying ground truth parameters are a diffusion coefficient of $0.03$ mm$^3$/day in gray matter and $0.3$ mm$^3$/day in white matter to account for preferential growth along fiber bundles in white matter \cite{g2012interstitial}, with a proliferation rate of $0.15$ day$^{-1}$ in the whole domain. These values are selected to be in biologically realistic ranges based on previous model calibration studies \cite{hormuth2021image}. The radiosensitivity parameter ratio is set to $\alpha/\beta = 10$ Gy following values reported in the literature \cite{rockne2009mathematical}. Additionally, we fix $\alpha_{\text{rt}} = 0.025$ Gy$^{-1}$ and $\alpha_{\text{ct}} = 0.9$. We assume $\beta_{\text{ct}} = 1.8$ hours for the clearance rate of the temozolomide chemotherapy agent \cite{wesolowski2010temozolomide}. The model Eq.~\eqref{eqn:rd-model} is discretized and solved as detailed in Section~\ref{sec:models}. To generate observations, the finite-element representation of the state is interpolated onto the native $T_1$ voxel image space and the resultant voxel measurements are polluted with 2\% Gaussian noise.

\subsection{Model verification}
\label{sec:model-verify}
To avoid an inverse crime \cite{kaipio2007statistical}, the inverse problem is solved on a coarser mesh with $137,261$ vertices. First order Lagrange elements are used for the state, parameter, and adjoint variables. Additionally, unlike the true parameter which depends on the underlying gray / white matter tissue, the modeled log diffusivity field does not explicitly incorporate this information. The inversion parameters $m_D, m_\kappa$ are modeled as Gaussian random fields as outlined in Section~\ref{sec:prior}. The hyper-parameters defining the prior are reported in Table~\ref{tab:sub-101-hyperparameters}.

\begin{table}[!htb]
\caption{Hyper-parameters for the Bayesian calibration of UPENN-GBM Subject 101.}
\centering
\begin{tabular}{cccc}
\hline
\multicolumn{4}{c}{Prior mean and variance of parameters}\\
\hline
\multicolumn{2}{c|}{$m_D$} & \multicolumn{2}{c}{$m_\kappa$} \\ 
\multicolumn{2}{c|}{$\log(mm^3/day)$} & \multicolumn{2}{c}{$\log(1/day)$} \\
Mean & \multicolumn{1}{c|}{Variance} & Mean & \multicolumn{1}{c}{Variance} \\
-1.30 & \multicolumn{1}{c|}{0.05} & -1.00 & \multicolumn{1}{c}{0.02} \\ \hline
\multicolumn{4}{c}{Spatial correlation lengths} \\ \hline
\multicolumn{2}{c|}{$\rho_{D}$ ($mm$)} & \multicolumn{2}{c}{$\rho_{\kappa}$ ($mm$)} \\
\multicolumn{2}{c|}{180} & \multicolumn{2}{c}{180} \\ \hline
\end{tabular}
\label{tab:sub-101-hyperparameters}
\end{table}

\begin{figure}
    \centering
    \includegraphics[width=\linewidth]{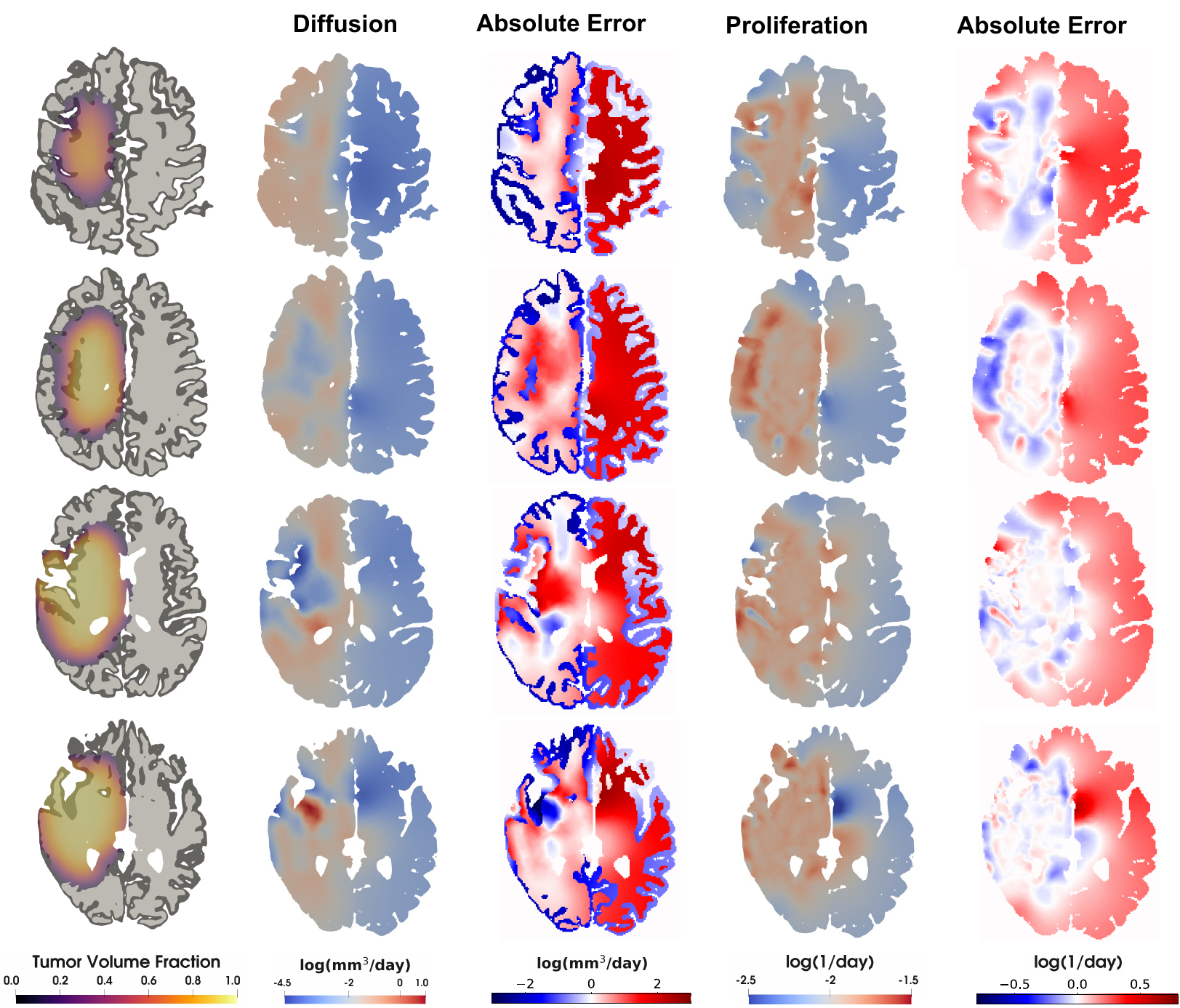}
    \caption{First column: axial slices showing white and gray matter segmentation with tumor state at final observation (week 12). Second column: \gls*{map} reconstruction of the log-diffusion field. Third column: absolute error in the reconstructed log-diffusion field. Fourth column: \gls*{map} reconstruction of the log-reaction field. Fifth column: absolute error in the reconstructed log-reaction field. Each log parameter field has discretized dimension $137,261$. Note the inferred heterogeneity in the reconstructed log-diffusion field as well as the spatial structure of the error where the parameters are well informed where the tumor is active.}
    \label{fig:map}
\end{figure}

The \gls*{map} reconstructions of the diffusion and proliferation coefficient fields for the daily imaging case are presented in Figure~\ref{fig:map}. In the reconstructed diffusion field, we observe that spatial heterogeneity arising from the underlying tissue type is captured. In the reconstructed proliferation rate, we do not see this effect, which is consistent with the true underlying parameter that is also homogeneous in the domain. Note in both cases that the \gls*{map} reconstruction is only well-informed where the tumor is active and the prior regularization dominates elsewhere. Furthermore, while the underlying structure in the true parameters is reconstructed, inference of the exact values appears challenging. The diffusion and reaction terms drive tumor dynamics in a complementary fashion, exacerbating the ill-posedness of the inverse problem and prompting more rigorous identifiability studies \cite{phillips2023assessing}. In spite of this, the reconstructed parameters lead to excellent reconstruction of the \glspl*{qoi}, as we will show in the next section.

\subsection{Assessing the value of additional imaging}
\label{sec:imgfreq-study}
\gls*{mri} data are highly informative, but expensive to collect and the resulting lack of information to assess response to therapy and progression may contribute to sub-optimal outcomes. One cannot accurately calibrate a model without data, nor can a clinician assess a patient's condition. Thus, there is a clear tension between the costs associated with additional imaging and the benefit of the additional information, for example, improving accuracy and reducing uncertainty in computational models used to predict patient prognosis. More accurate and precise models would offer better insights when tailoring interventions. We seek to better understand the effect of data availability on both parameter estimation and the prediction of clinically relevant \glspl*{qoi}.

\begin{figure}[!htb]
    \centering
    \includegraphics[width=0.8\textwidth]{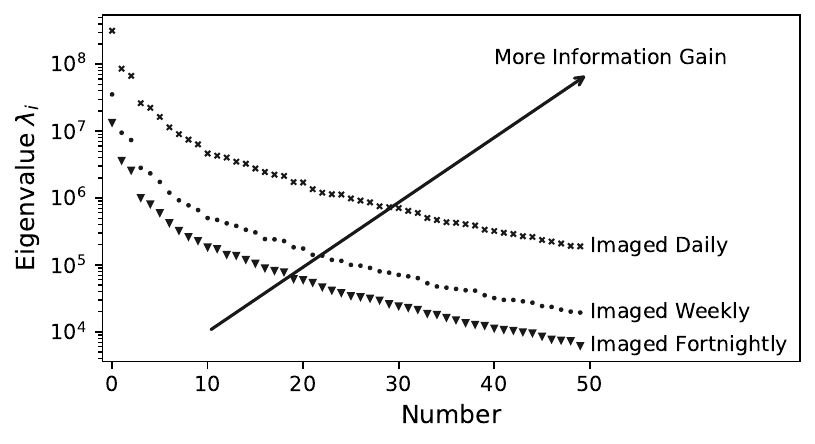}
    \caption{Spectral decay of the prior-preconditioned Hessian for various imaging frequencies. The larger eigenvalues associated with more frequent imaging indicate more information gain in the corresponding eigenvectors.}
    \label{fig:spectral_decay}
\end{figure}

Recall that Eq.~\eqref{eqn:covpost} expresses the posterior covariance as a low-rank update to the prior covariance through the prior-preconditioned eigenvectors of the data-misfit Hessian given by Eq.~\eqref{eqn:GEVP}. The associated eigenvalues quantify how much information is extracted from the data, while the spatial structure of the eigenvectors sheds insight into the directions in parameter space which are informed by the data. The spectrum of the prior-preconditioned Hessian, defined by Eq.~\eqref{eqn:GEVP}, is computed for the first $50$ eigenvalues and plotted for the three imaging frequency scenarios in Figure~\ref{fig:spectral_decay}. As expected, the magnitude of eigenvalues is larger when the patient is observed more frequently, indicating larger corrections to the prior covariance when calculating the low-rank approximation to the posterior covariance \eqref{eqn:lr-covpost}. In other words, the directions in parameter space represented by the associated eigenvectors are better informed with more data.

\begin{figure}[!htb]
    \centering
    \includegraphics[width=\linewidth]{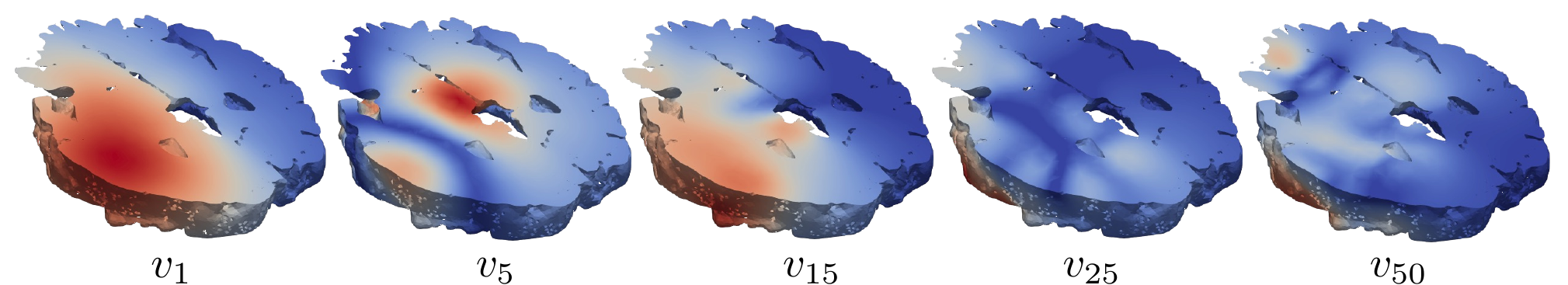}
    \caption{Prior-preconditioned eigenvectors of the data-misfit Hessian, $\mathbf{H_{\text{misfit}}}$, for UPENN-GBM subject 101 when imaged daily. The different eigenvectors localize different regions of the brain, with the most informative modes concentrated in regions with active tumor growth. Eigenvectors corresponding to smaller eigenvalues are increasingly oscillatory, resolving fine scale dynamics while also being less informed by the data.}
    \label{fig:sub-00101-evec}
\end{figure}

Figure~\ref{fig:sub-00101-evec} displays several of the associated eigenvectors, when imaged daily. The localization of meaningful information observed in the \gls*{map} is again evident in the eigenvectors corresponding to the largest eigenvalues; that is, the modes most informed by the data. Additionally, we can see that eigenvectors corresponding to smaller eigenvalues capture higher frequency content from the data, though they do not impart as large of a correction to the prior covariance.

\begin{figure}[!htb]
    \centering
    \subfloat[]{\includegraphics[width=0.49\linewidth]{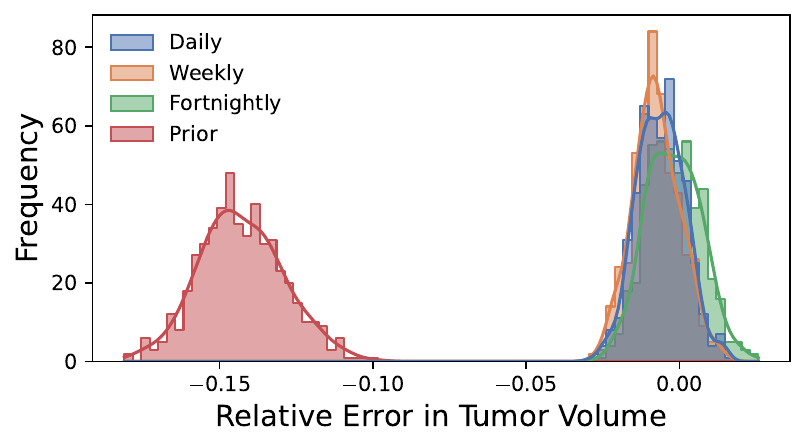}}
    \subfloat[]{\includegraphics[width=0.49\linewidth]{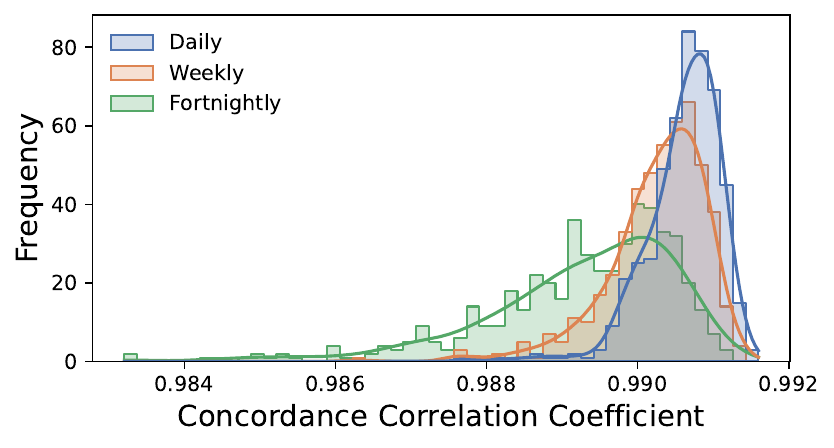}}
    \caption{Predictive distributions of (a) relative error in total tumor volume and (b) concordance correlation coefficient for various imaging schedules. The prior performs significantly ($p<0.001$) worse in both \glspl*{qoi} than the calibrated models and is omitted from (b) to focus on the relative performance. Overall, calibration to patient data improves model predictive quality and significantly reduces variance ($p<0.001$) in predictive distributions when compared to sampling from the prior distribution. All calibrated models capture the general tumor growth as measured by the volume, while the concordance correlation coefficient shows model accuracy increasing with additional observational data.}
    \label{fig:sub-00101-pushforward}%
\end{figure}

We assess the impact on the model's predictive performance by computing two quantities of interest: the relative error in total tumor volume and concordance correlation coefficient, calculated as outlined in Section~\ref{sec:qoi}. The pushforward is computed by drawing $500$ Monte Carlo samples from the prior, $\boldsymbol{m} \sim \nu_{\text{pr}}$, as well as the Laplace approximation to the posterior for the three cases, $\boldsymbol{m} \sim \nu^{\text{LA}}_{\text{post}}$. The sampled parameters are then used to simulate tumor growth from the end of the imaging to the one month prediction; i.e., we compute the pushforwards $(q_{\text{TV}}\circ \mathcal{F}_{\text{pred}})_\sharp \nu$ and $(q_{\text{CCC}}\circ \mathcal{F}_{\text{pred}})_\sharp \nu$, with $\nu=\nu_{\text{pr}}$ or $\nu_{\text{post}}^{\text{LA}}$. The pushforward distributions are reported in Figure~\ref{fig:sub-00101-pushforward} for the two \glspl*{qoi}.

We observe that the predictive performance of the model improves in both \glspl*{qoi} as the patient is imaged more frequently and all models calibrated with patient-specific data significantly ($p<0.001$) outperform the prior as determined by a Mann-Whitney U-test \cite{sachs2012applied}. As anticipated, more information leads to better prediction of tumor status and significantly reduced variance in both \glspl*{qoi} ($p<0.001$) in the predictive distributions compared to the prior as determined with Levene's test \cite{sachs2012applied}. For this particular study, the prior distribution underpredicts the true tumor volume while all calibrations capture the true value. The \gls*{ccc} indicates an ordering between the calibrated models, with additional imaging improving performance, though there is a diminishing return as was observed in Figure~\ref{fig:spectral_decay}.

These results illustrate that there is a tradeoff between information gain and imaging frequency. An exciting direction for future work would be to formulate an optimal experimental design problem to optimally choose the imaging frequency balancing improved model predictive quality with the cost of imaging. Related approaches have shown promise for one-dimensional radiotherapy planning \cite{cho2020bayesian,cho2023adaptive}, however, in the \gls*{pde} setting reduced order modeling \cite{benner2015survey} and multifidelity methods \cite{peherstorfer2018survey} will be crucial to ensure tractability of the resultant optimization problem. Moreover, coupling the experimental design question to therapy optimization \cite{fuentes2010adaptive,fahrenholtz2015model,colli2021optimal} will be crucial for assessing potential clinical impact. In the presence of uncertainty, this will require characterization of risk and optimization of statistical properties of the \glspl*{qoi} \cite{kouri2016risk}.

\subsection{Enabling computational tractability in clinically relevant timeframes}
\label{sec:implementation}
It is crucial to deliver insights in a clinically relevant timeframe. While the methods developed in Section~\ref{sec:la-post} scale with the intrinsic difficulty of the inverse problem rather than the discretization, the computational burden of the nonlinear forward \gls*{pde} solve and gradient and Hessian actions is still of concern. We need an efficient, parallelized implementation of the forward model to complement the algorithmic scalability. To this end, we interface with \texttt{PETSc} \cite{dalcin2011parallel} to extend \texttt{hIPPYlib} \cite{villa2021hippylib} and \texttt{FEniCS} \cite{alnaes2015fenics} to enable fast, distributed solution on high-performance computing resources. In particular, by interfacing directly with \texttt{PETSc}, one gains access to high-quality solvers and a wide variety of preconditioners useful for fast solution of the forward model; for example, we utilize \texttt{BoomerAMG} \cite{yang2002boomeramg}. Futhermore, implementation of the forward model in \texttt{FEniCS} reduces the overhead associated with exploring models other than the reaction-diffusion Eq.~\eqref{eqn:rd-model} to the ability to write the variational form.

\begin{figure}[!htb]
    \centering
    \subfloat[Intra-node]{\includegraphics[width=0.49\linewidth]{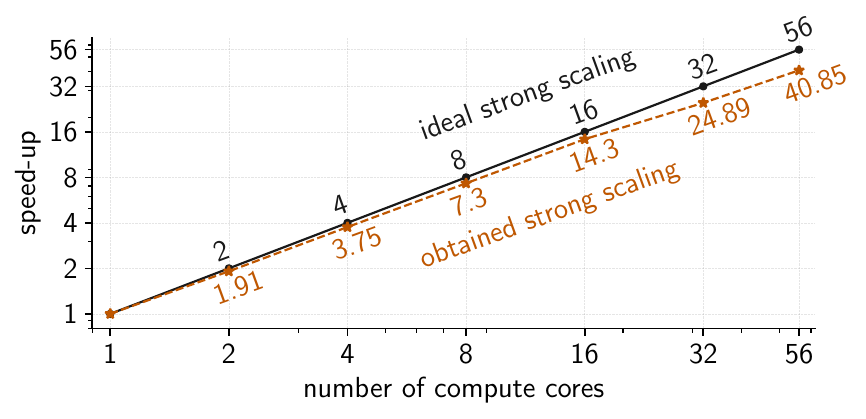}}
    \subfloat[Inter-node]{\includegraphics[width=0.49\linewidth]{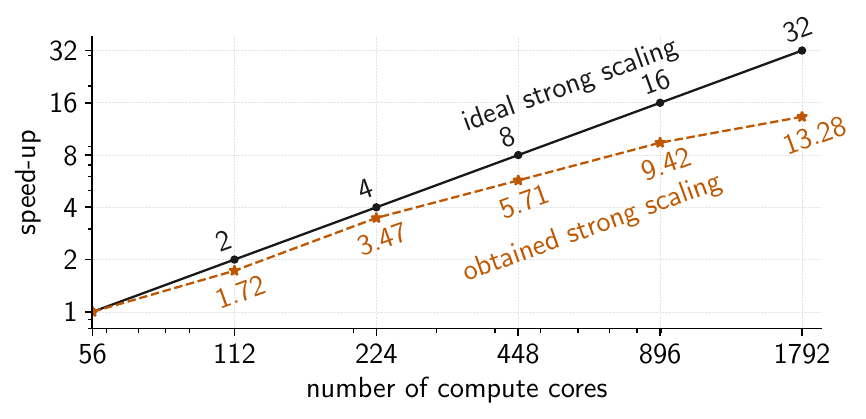}}
    \caption{Strong scaling of the forward solve obtaining a one-month prediction in (a) 42.2 seconds on one node for a benchmark problem with 1 million \glspl*{dof} and (b) 12.2 seconds on 32 nodes for a benchmark problem with 4 million \glspl*{dof}.}
    \label{fig:strong-scaling}%
\end{figure}

Scalability studies are performed on the Frontera supercomputer at the \gls*{tacc} \cite{stanzione2020frontera} and are reported in Figure~\ref{fig:strong-scaling}. For the scalability study, a tumor seed defined by a Gaussian function is grown within a $100$ mm$^3$ domain with 1 million \glspl*{dof} (for the intra-node case) and 4 million \glspl*{dof} (for the inter-node case). First-order Lagrange elements are used for the state, parameter, and adjoint variable. An implicit Euler discretization is used in time with a step size of one day. We achieve near perfect strong scaling within node as well as good scaling across nodes up to 32 nodes. Similar performance is achieved for the linear adjoint solves and is reported in \ref{sec:appendix-scaling}. Strong scalability is of the greatest importance in this context as we wish to accelerate computation on a pre-defined, fixed computational domain. However, a weak-scaling study was also performed and demonstrated adequate scalability, with results reported in \ref{sec:appendix-scaling}. The code for the forward and inverse solvers is publicly available at: \hyperlink{https://github.com/gtpash/dt4co}{https://github.com/gtpash/dt4co}.

%% file: sections/ivygap_case_study.tex
\section{Model validation and uncertainty quantification in the clinical setting}
\label{sec:ivygap}

Model validation studies play a key role in establishing the clinical relevance of predictive science and certifying the computational models underpinning a digital twin. We further demonstrate the high-dimensional Bayesian model calibration process on a subset of patients with historical longitudinal \gls*{mri} data from the publicly available IvyGAP dataset \cite{puchalski2018anatomic,shah2016tcia}. We first detail the data and experimental setup. Performance across the cohort is then assessed by computing pushforward distributions for relevant \glspl*{qoi} in a variety of prediction settings. Finally, we address the role of model inadequacy and give an outlook.

\subsection{Experimental setup and Bayesian model calibration}
\label{sec:ivygap_experiment}
\begin{figure}[!htb]
    \centering
    \includegraphics[width=\linewidth]{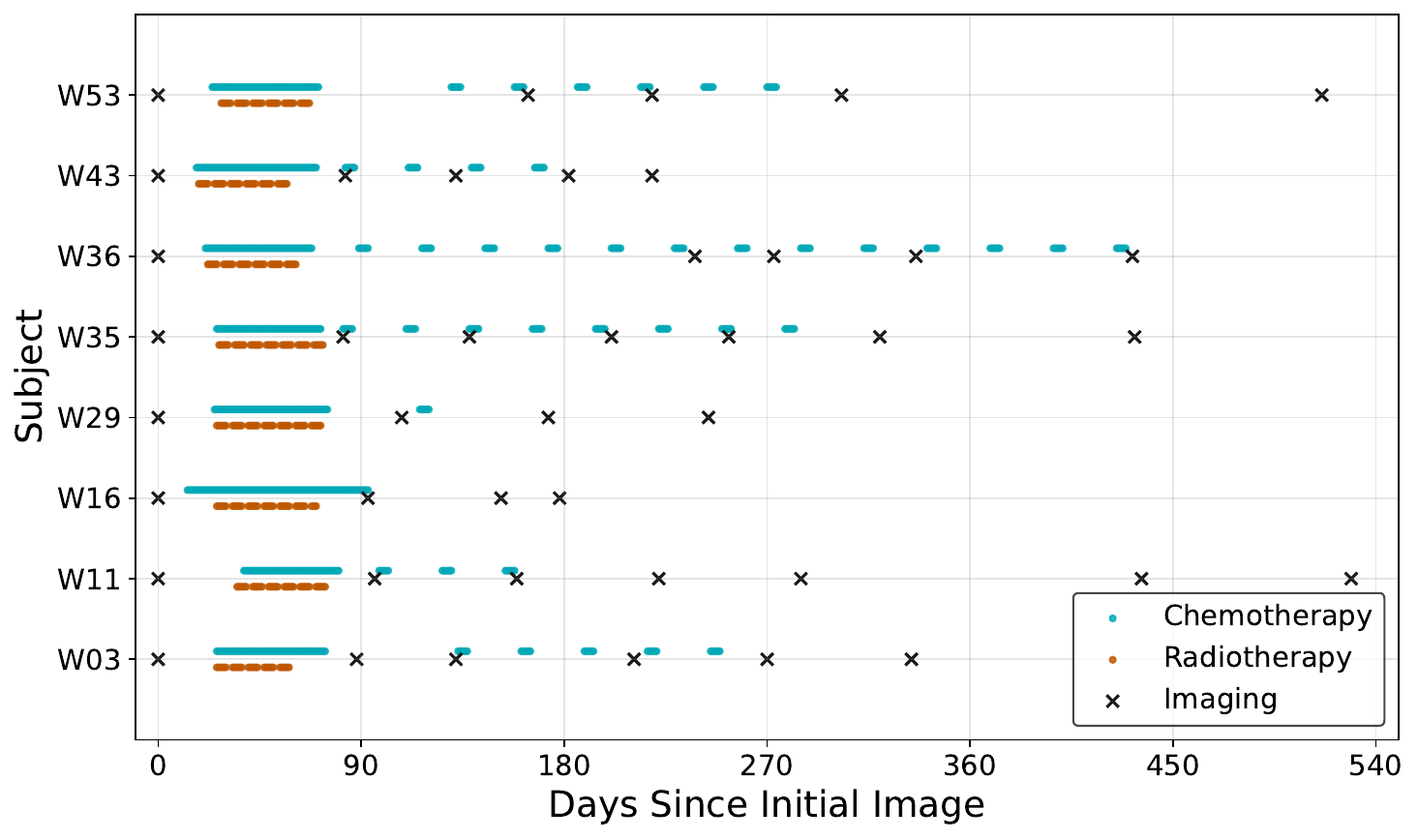}
    \caption{Imaging and treatment timelines for the IvyGAP cohort used for the model validation study. There is considerable variation amongst the patients in the number and frequency of data collection as well as the treatment schedules received.}
    \label{fig:cohort-timelines}
\end{figure}

The IvyGAP dataset contains \gls*{mri} data (including the types discussed in Section~\ref{sec:data}) acquired during the course of the patient's treatment. Additionally, the dataset contains information about the radiotherapy and chemotherapy treatment schedules that the patients received. As the IvyGAP dataset does not contain tumor segmentations, the ONCOhabitats \cite{juan2019oncohabitats} tool is used to generate an initial segmentation of the enhancing and non-enhancing tumor regions. These regions are then manually corrected. Tumor cellularity and volume fraction are computed as in Section~\ref{sec:data}. In all results, each patient's tumor state is estimated from data at the first imaging time, $d(\bar{x}, t_0)$. If there was a resection event, we use the first post-resection image in the dataset as the baseline. Additionally, we restrict the computational domain to only the hemisphere containing the lesion since all of the tumors considered in this study are unifocal and do not invade the contralateral hemisphere. This results in meshes with approximately $250,000$ vertices for all cases in this study. The imaging and treatment timelines for the cohort are shown in Figure~\ref{fig:cohort-timelines}. We observe a wide variation  in both imaging schedules and treatment regimens.

We infer the log diffusivity field, $m_D$, and the log proliferation rate, $m_\kappa$, of the reaction-diffusion model Eq.~\eqref{eqn:rd-model}. Additionally, the log diffusivity field is assumed to be uncorrelated across the gray and white matter interface $m_D = m_{D,gm} \chi_{gm} + m_{D,wm} (1-\chi_{gm})$ where the gray matter indicator function $\chi_{gm}$ is obtained from the tissue segmentation derived in Section~\ref{sec:data}. Each log parameter is modeled using the Gaussian random field prior detailed in Sec.~\ref{sec:prior}. To determine the prior mean and variance for each of the parameters, an initial calibration of the cohort is performed where the parameters are modeled as scalar quantities. The correlation length is set to approximately the largest dimension of the domain, with larger correlation lengths allowed in white matter. The variance of the noise is taken to be equivalent to $6.25\%$ noise, following the reported value in \cite{liang2023bayesian} for a similar data-acquisition and tumor volume fraction estimation pipeline. The hyper-parameters defining the priors are summarized in Table~\ref{tab:cohort-prior}.

\begin{table}[!htb]
\caption{Estimated hyper-parameters for the Bayesian calibration.}
\centering
\begin{tabular}{cccccccc}
\hline
\multicolumn{8}{c}{Prior mean and variance of parameters}\\
\hline
\multicolumn{2}{c|}{\begin{tabular}[c]{@{}c@{}}$m_\kappa$\\$\log(1/day)$\end{tabular}} & \multicolumn{2}{c|}{\begin{tabular}[c]{@{}c@{}}$m_D$\\ $\log(mm^3/day)$\end{tabular}} & \multicolumn{2}{c|}{\begin{tabular}[c]{@{}c@{}}$m_{D,wm})$\\ $\log(mm^3/day)$\end{tabular}} & \multicolumn{2}{c}{\begin{tabular}[c]{@{}c@{}}$m_{D,gm}$\\ $\log(mm^3/day)$\end{tabular}} \\
Mean & \multicolumn{1}{c|}{Variance} & Mean                                & \multicolumn{1}{c|}{Variance} & Mean & \multicolumn{1}{c|}{Variance} & Mean & Variance \\
-1.230 & \multicolumn{1}{c|}{0.040} & -1.167 & \multicolumn{1}{c|}{0.115} & -0.991 & \multicolumn{1}{c|}{0.115} & -1.467 & 0.115 \\ \hline
\multicolumn{8}{c}{Spatial correlation lengths and noise variance} \\ \hline
\multicolumn{2}{c|}{$\rho_{\kappa}$ ($mm$)} & \multicolumn{2}{c|}{$\rho_{D,gm}$ ($mm$)} & \multicolumn{2}{c|}{$\rho_{D,wm}$ ($mm$)} & \multicolumn{2}{c}{$\sigma_{noise}^2$} \\
\multicolumn{2}{c|}{180} & \multicolumn{2}{c|}{180} & \multicolumn{2}{c|}{360} & \multicolumn{2}{c}{3.9e-3} \\ \hline
\end{tabular}
\label{tab:cohort-prior}
\end{table}

For model validation, the last image in the dataset is withheld from the calibration and is set aside to serve as a prediction target. Numerical experiments were performed on four nodes of \gls*{tacc}'s Frontera supercomputer. An implicit Euler discretization was used in time, with a time step of one day. The chemoradiation model Eq.~\eqref{eqn:treatment-model} is fixed for all patients and implemented as in Section~\ref{sec:upenn}, with the exception that $\alpha_{\text{ct}} = 0.82$ to match the calibrated therapy effects reported in \cite{hormuth2021image}. A maximum of $50$ Newton iterations are used to compute the \gls*{map} point by solving the nonlinear optimization problem Eq.~\eqref{eqn:map} using the relative decrease in the norm of the gradient as a convergence criterion. The required time to compute the \gls*{map} point ranged extensively due to the varied lengths of the patient imaging timelines and the number of required inner \gls*{cg} iterations, with a median wall clock time of $14.2$ hours (IQR $10.4$-$18.3$ hours). The low-rank approximation of the posterior covariance $\mathbf{\Gamma}_{\text{post}}$ given by Eq.~\eqref{eqn:covpost} was computed with $r=50$ and an oversampling factor of $10$ for the randomized eigensolver and took approximately three hours in all cases. 

\subsection{Predictive performance on cohort}
\label{sec:ivygap_performance}
\begin{figure}[!htb]
    \centering
    \subfloat[]{\includegraphics[width=0.49\linewidth]{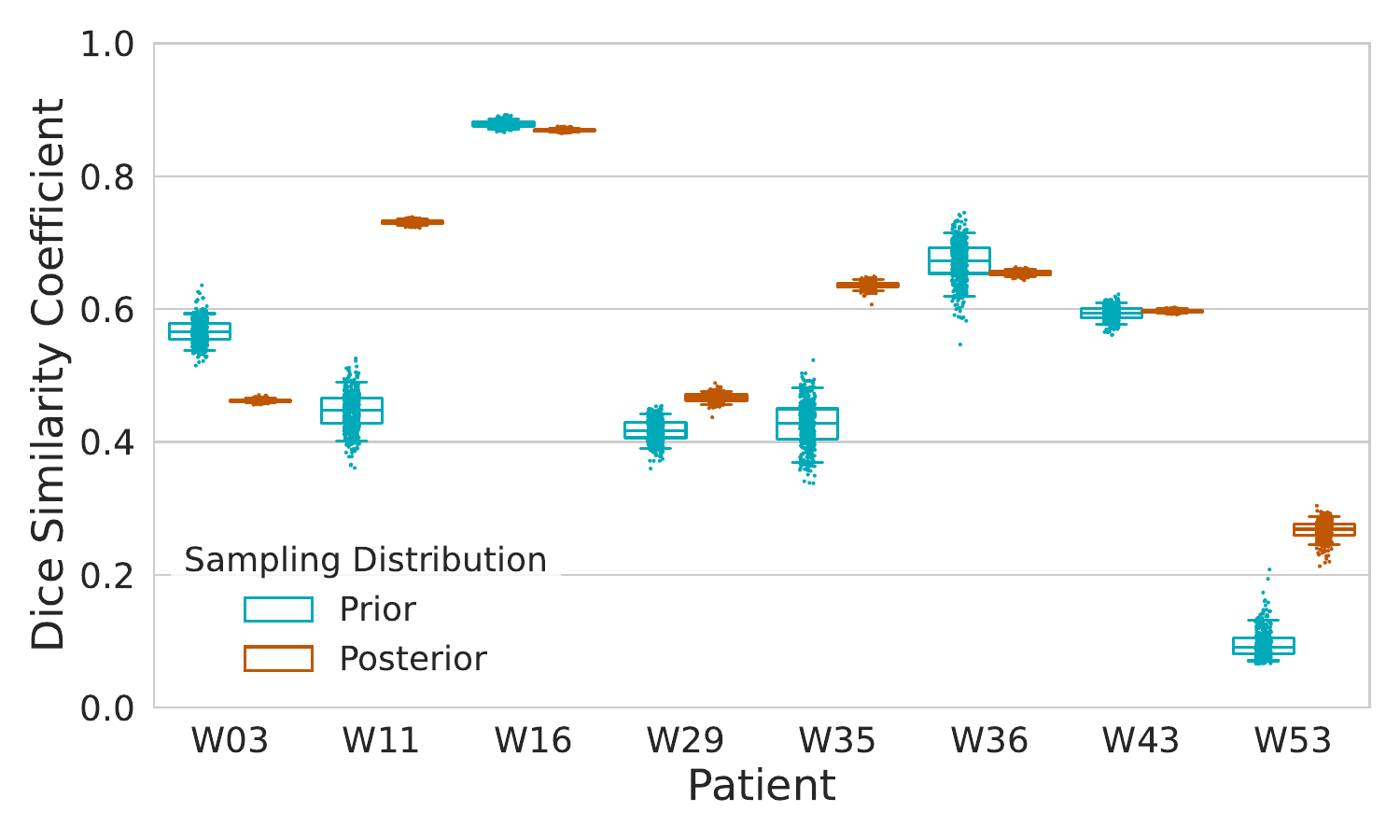}}
    \subfloat[]{\includegraphics[width=0.49\linewidth]{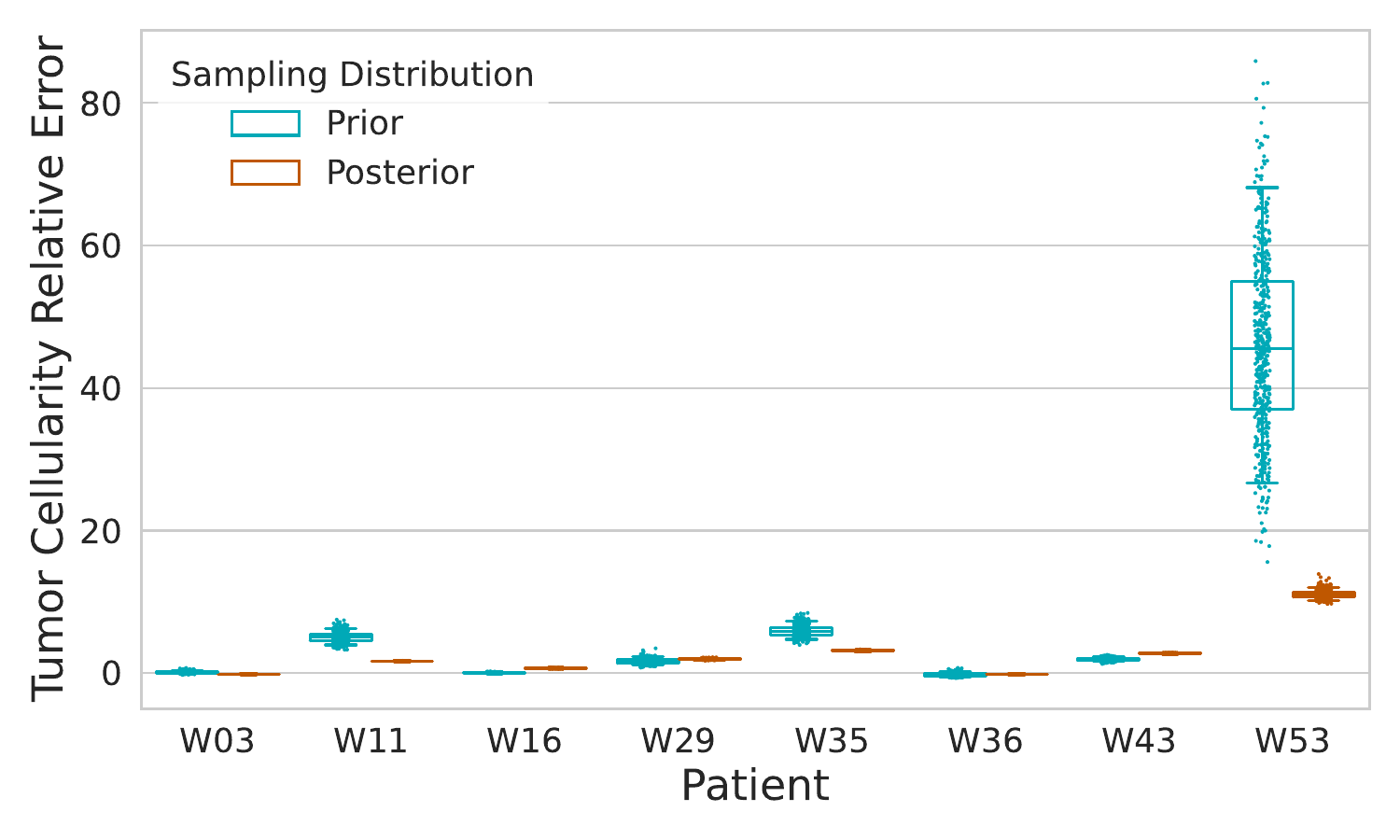}}
    \caption{Prior and posterior predictive distributions of (a) Dice similarity coefficient and (b) total tumor cellularity for the IvyGAP patient cohort. The simulation window is from the last scan used for model calibration to the last acquired image.}
    \label{fig:l2f-pushforward}%
\end{figure}

Model performance is assessed on the cohort by generating posterior predictive distributions of \glspl*{qoi}. In particular, the Dice similarity coefficient and relative error in total tumor cellularity are used and computed as described in Section~\ref{sec:qoi}. To mimic the envisioned clinical deployment, the initial condition is taken to be the penultimate scan (the last data seen during model calibration) and the inferred parameters are used to make a prediction for comparison at the time last scan. That is, $u_0$ is determined from $d(\bar{x},t_{n_t-1})$ and the simulation window is $(t_{n_t-1}, t_f)$ with $t_{n_t-1}$ denoting the time the second to last measurement. Figure~\ref{fig:l2f-pushforward} reports the predictive distributions, with $500$ parameter samples from both the prior distribution and the low-rank based Laplace approximation to the posterior. Summary statistics are tabulated and reported in \ref{sec:appendix-summary-statistics}. The reduced variance of the low-rank Laplace approximations is evident across both cohorts and across the both \glspl*{qoi}. While not all patients see a clear separation from the prior, such as W11 or W35, the posterior predictive distributions based on the Laplace approximation typically exhibit better spatial agreement with the observed tumor extent, as measured by the Dice coefficient. Additionally, the posterior predictive distributions appear to capture the total tumor cellularity well. Overall, the subject-specific calibrated model demonstrate a much improved ability to capture tumor progression.

\begin{figure}[!htb]
    \centering
    \subfloat[]{\includegraphics[width=0.49\linewidth]{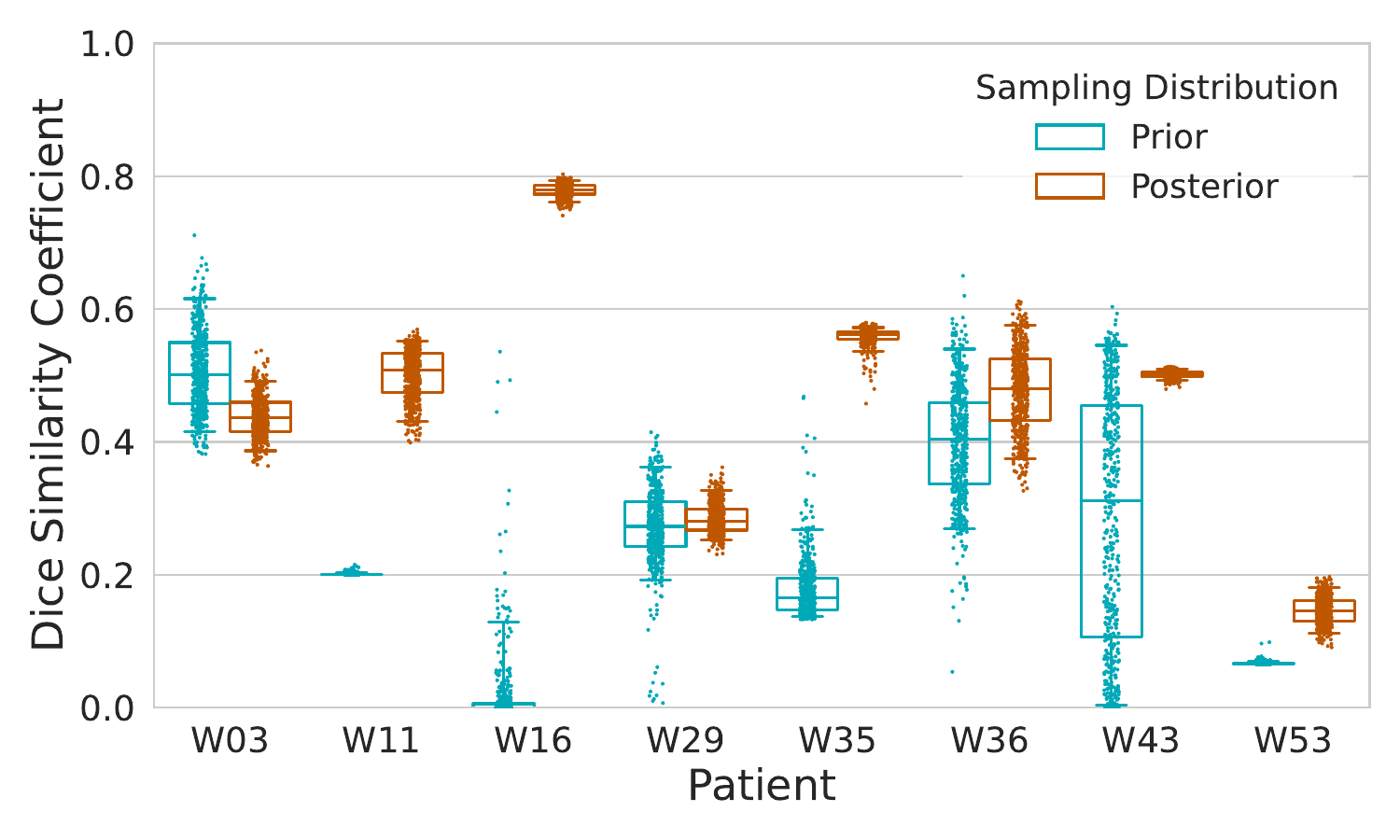}}
    \subfloat[]{\includegraphics[width=0.49\linewidth]{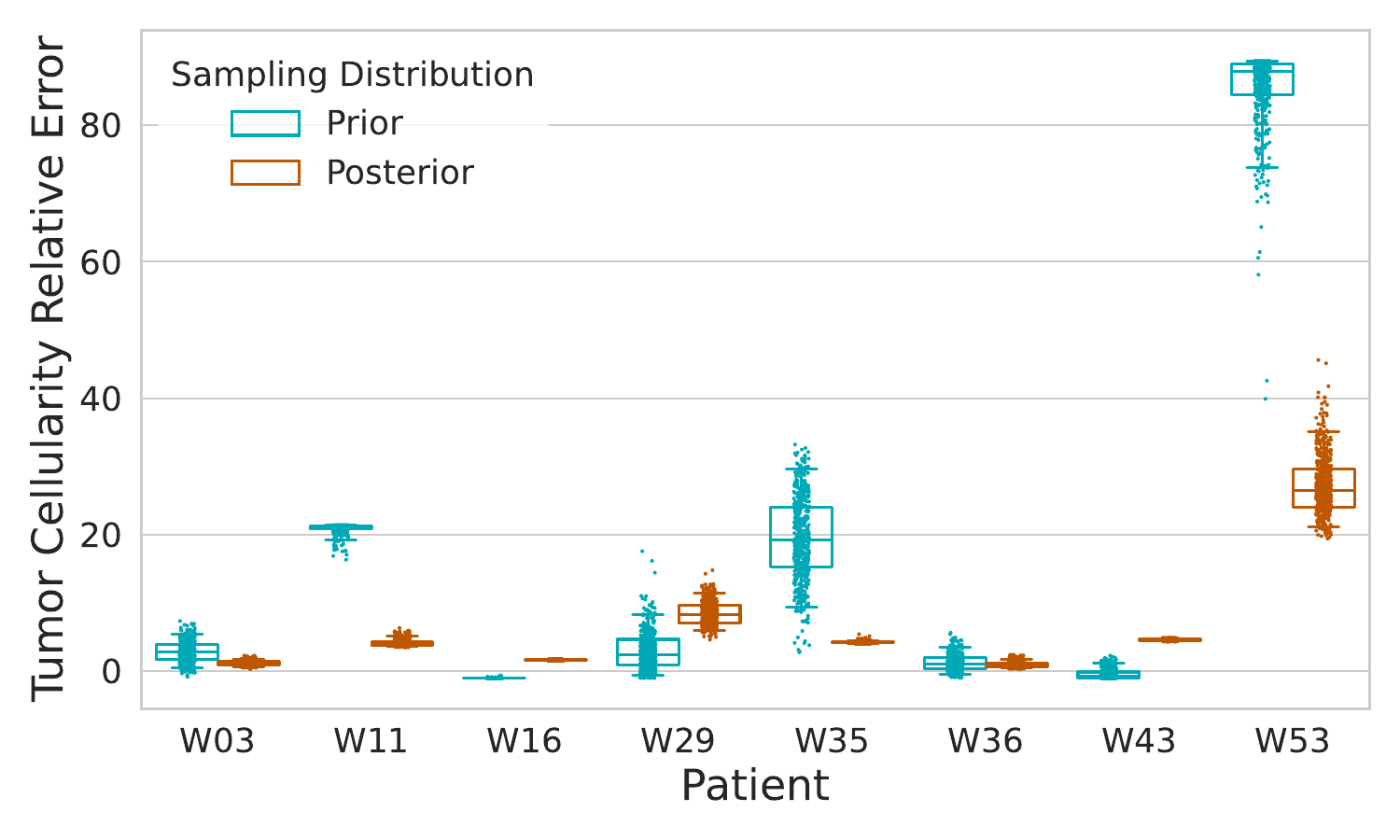}}
    \caption{Posterior predictive distributions of (a) Dice similarity coefficient and (b) total tumor cellularity for the IvyGAP patient cohort. The simulation window is from the first acquired scan until the last acquired image, through the calibration window.}
    \label{fig:i2f-pushforward}%
\end{figure}

To assess the stability of the inferred parameters, we consider the case where prediction is made from the first visit through the calibration window to the final observation. The prior and posterior pushforward distributions are presented in Figure~\ref{fig:i2f-pushforward}. As before, summary statistics are reported in \ref{sec:appendix-summary-statistics}. While tumor cores are unlikely to change over short horizons, the longer simulation window exacerbates accumulated errors as the model diverges from the observed data, and manifests in increased variance in the pushforwards. This also highlights structural deficiencies in the models calibrated to intermediate observations. We observe that in many cases, there is only a slight drop in the predictive performance of the posteriors over this longer horizon and conclude that the calibration does in fact provide a robust prediction of both the tumor shape (as measured by the Dice coefficient) and the intra-tumoral heterogeneity (as measured by the total tumor cellularity). This is in stark contrast to the prior distribution, where model inadequacy compounds over the long prediction horizon manifesting in large variability and drop in predictive performance. 

\subsection{Model inadequacy and opportunities}
\begin{figure}[!htb]
    \centering
    \includegraphics[width=\linewidth]{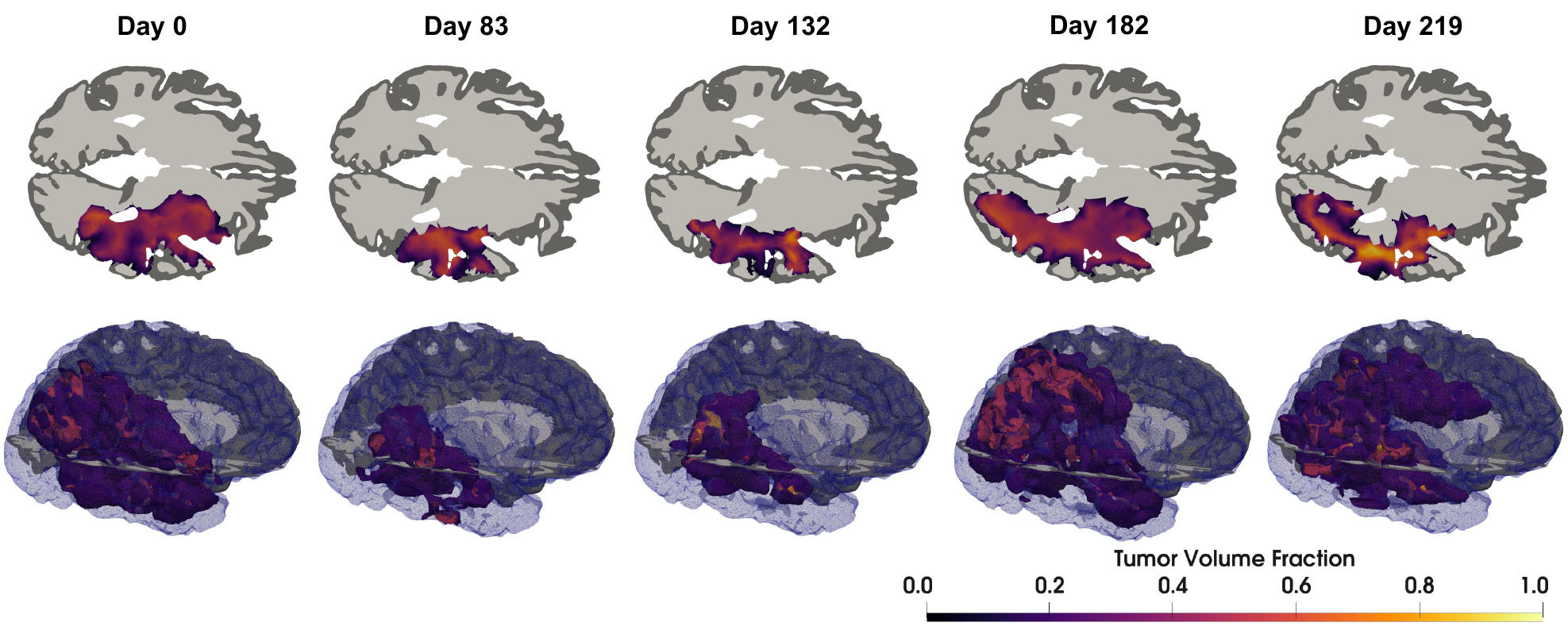}
    \caption{Observed disease progression of IvyGAP patient W43 throughout the course of treatment. The top row is an axial cross-sectional view near the middle of the tumor. The second row is a volumetric rendering of the computational domain, tumor, and axial slice. Observe the strong effect of chemoradiation treatment in the first three snapshots. In the final two snapshots, note the apparent tumor retreat in some areas despite the lack of therapy, while there is aggressive invasion in other regions.}
    \label{fig:sub-W43-progression}
\end{figure}

The predictive performance across the cohort was shown to be both stable over long time horizons and meaningfully informed by the data, especially given the sparse data collection. However, the model is far from a perfect match with reality. To better understand the limitations of the model and calibration, we focus on subject W43 as a representative patient. The patient's disease progression is visualized in Figure~\ref{fig:sub-W43-progression}. The drastic change in the tumor extent during the course of therapy, and especially between the last two snapshots clearly poses a significant modeling challenge. In particular, there are two phenomena that are difficult for the formulated model to capture: (1) the apparent retreat of the tumor in other regions while not under active treatment and (2) the rapid progression of disease into new regions.

Change in the non-enhancing tumor region is one biological explanation for the apparent retreat of the tumor between the final two snapshots. The non-enhancing tumor region is comprised of both edema and infiltration and it is notoriously difficult to accurately estimate the tumor cellularity \cite{puchalski2018anatomic}. For instance, changes in steroid dose and schedule can impact the presentation of the non-enhancing region of the tumor. Since the enhancing and non-enhancing regions are combined to estimate the tumor \gls*{roi}, large variations may appear as aggressive invasion or rapid retreat in the data. The reaction-diffusion model employed may fail to capture these fluctuations in the non-enhancing tumor component since it cannot adequately resolve the complex dynamics related to treatment and biology in this region. Alternative imaging modalities that better resolve intra-tumoral heterogeneity should be considered to improve data fidelity \cite{hu2020imaging}.

\begin{figure}[!htb]
    \centering
    \includegraphics[width=\linewidth]{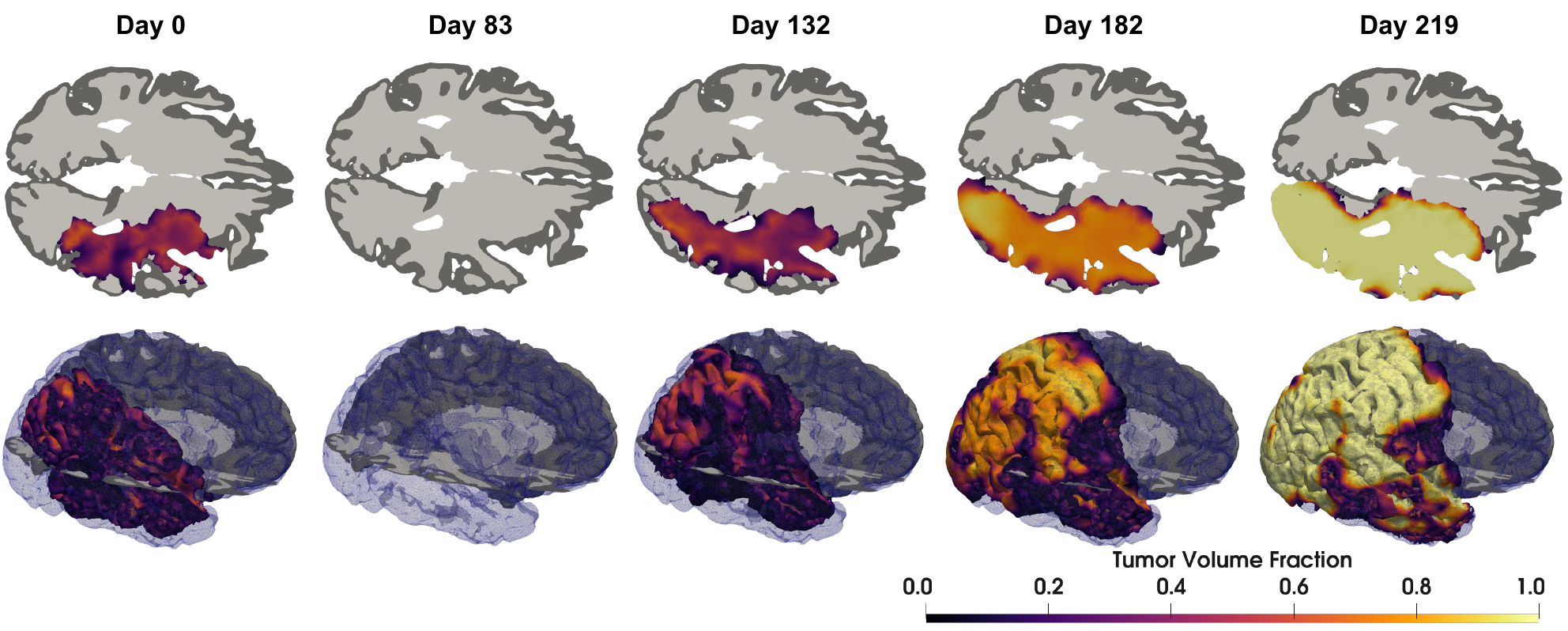}
    \caption{Predicted tumor progression from the first visit to the last visit for patient W43 using the \gls*{map} point. The top row is an axial cross-sectional view near the middle of the tumor. The second row is a volumetric rendering of the computational domain, tumor, and axial slice. The model overpredicts response to therapy compared to the observed data in this case (c.f. Figure \ref{fig:sub-W43-progression}). Additionally, the model overpredicts the recurrence and struggles to capture internal heterogeneity of the tumor at the prediction time.}
    \label{fig:sub-W43-i2f-progression}
\end{figure}

Furthermore, the fixed treatment model is not appropriate for every patient and contributes to model inadequacy. Consider the initial-to-final prediction case. The \gls*{map} point prediction of the tumor progression is presented in Figure~\ref{fig:sub-W43-i2f-progression}. For this specific patient, we observe that the specified chemoradiation therapy model is too strong, leading to an unrealistic near total predicted remission at the second visit. Patient-specific calibration of treatment models may improve predictive quality, but must be balanced with identifiability of the model parameters given the dearth of observational data.

\begin{figure}[!htb]
    \centering
    \includegraphics[width=\linewidth]{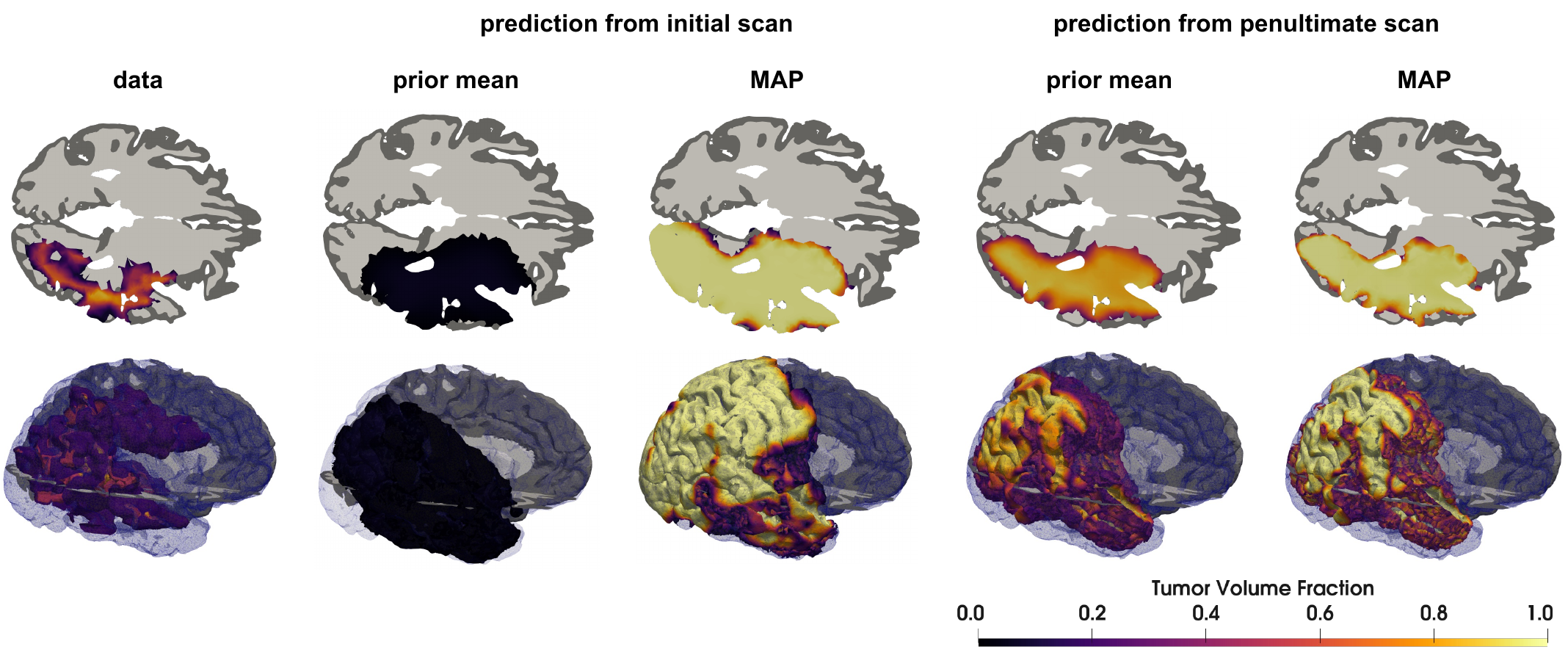}
    \caption{Predicted final tumor state for IvyGAP patient W43 using the \gls*{map} point and prior mean both from the initial and penultimate scans as initial condition. The top row is an axial cross-sectional view near the middle of the tumor. The second row is a volumetric rendering of the computational domain, tumor, and axial slice. The prior mean generates a poor prediction from the initial scan indicating that it is not appropriate for this patient. The short prediction horizon from the penultimate scan leads to qualitatively similar structure between the prior and calibrated models. The calibrated models generate qualitatively similar predictions both over the long and short prediction horizons.}
    \label{fig:W43-pred}
\end{figure}

While the total tumor cellularity results in Figure~\ref{fig:l2f-pushforward} indicate potentially better performance of the calibrated model to predict the internal growth dynamics of the tumor core, both the \gls*{map} and prior mean generate qualitatively similar predictions of tumor shape and extent for the last-to-final case, as shown in Figure~\ref{fig:W43-pred}. This highlights the challenging nature of extrapolation and motivates further development of predictive models to better capture these effects. One approach may be to generate patient-specific priors based on genetic subtype as recurrent gliomas typically present with similar molecular structure \cite{barthel2019longitudinal} and epigenetic classification has demonstrated prognostic value \cite{ceccarelli2016molecular}. The reaction-diffusion model of tumor growth combines multiple tumor development mechanisms into its two biophysical parameters and is certainly inadequate to capture the full heterogeneity present. More complex models are necessary to improve accuracy, incorporating mass effect \cite{hogea2007modeling,hormuth2018mechanically}, multi-species  \cite{hormuth2021image,fritz2021analysis}, metabolic processes \cite{laubenbacher2009systems,clarke2020executable}, or vascular structure \cite{hormuth2019calibrating,fritz2021modeling}, amongst others. Many of these models can readily be brought into the presented framework. Furthermore, development and calibration of more sophisticated models for radio- and chemo-therapy will be necessary to increase fidelity for optimization of therapeutic regimens.

%% file: sections/conclusion.tex
\section{Conclusion}
\label{sec:conclusion}
This work developed an end-to-end Bayesian framework for the integration of \gls*{mri} data with mathematical models of tumor growth to enable digital twins in precision oncology. High-fidelity computational representations of the patient anatomy are generated from \gls*{mri} data. Patient-specific biophysical model parameters are inferred from longitudinally collected \gls*{mri} measurements and account for uncertainty in the data acquisition and processing. The spatially varying model parameters are high-dimensional upon discretization and pose a significant computational challenge. To ensure tractability of the inverse problem, an efficient parallel implementation of the forward model was coupled with a scalable adjoint-based Newton-Krylov optimization algorithm. To overcome the prohibitive computational cost of performing \gls*{mcmc} to explore the posterior distribution, we employ the Laplace approximation to the posterior and make use of a low-rank approximation for rapid sampling. The Bayesian approach rigorously quantifies uncertainties in the inferred parameters, which are then propagated to predict future tumor growth and key clinical quantities of interest. The broad utility of the proposed approach was demonstrated by application to patients from two publicly available datasets. Moreover, the flexible implementation of the framework could readily be translated to other other organs, such as the breast \cite{jarrett2021quantitative} and prostate \cite{lorenzo2019computer}. To the best of our knowledge, this is the first development and demonstration of such an end-to-end Bayesian pipeline on clinical data accounting for the complex anatomy of the human brain.

The methodology was verified to accurately capture intra-tumoral spatial heterogeneity in an \textit{in silico} case study where the underlying growth mechanism was controlled. Patient-specific posteriors were shown to improve model predictive quality of key clinical metrics, while reducing variance in the model parameters. Alternate imaging schedules were implemented and used to quantitatively assess the information gain and value of additional imaging. Given the significant cost of \gls*{mri}, these results motivate an interesting optimal experimental design question regarding how frequently \gls*{mri} data should be collected as well as \textit{when} patients should be imaged. To assess the reliability of the approach in the clinic, the model was calibrated to a cohort of patients using publicly available clinical data. The posterior predictive distributions demonstrated robust agreement both of tumor shape, as measured by the Dice coefficient, and structure, as measured by the relative error in total tumor cellularity. This validation study also revealed future directions for model improvement and development to address key questions of model inadequacy to improve realism and predictive power. Together, these two studies establish the capacity of predictive science in oncology to provide a strong foundation for the development of digital twins.

\section{Acknowledgments}
\label{sec:Acknowledgments}
The authors thank Michael Kapteyn for insightful comments on an earlier draft of this work.

GP acknowledges support from the Office of Science, Advanced Scientific Computing Research, U.S. Department of Energy Computational Science Graduate Fellowship under Award Number DE-SC0021110. GP and KW acknowledge support from Department of Energy grant DE-SC002317, the National Science Foundation (NSF) grant 2436499, and NSF FDT-Biotech award 2436499. DAH acknowledges funding from CPRIT RP220225, NSF DMS 2436499, and the American Cancer Society IRG-21-135-01-IRG. TEY acknowledges funding from the National Cancer Institute R01CA235800, U24CA226110, U01CA174706. TEY is a CPRIT Scholar in Cancer Research. The authors acknowledge the Texas Advanced Computing Center (TACC) at The University of Texas at Austin for providing computational resources (Frontera) that have contributed to the research results reported within this paper. URL: \href{https://tacc.utexas.edu/}{http://www.tacc.utexas.edu}

This report was prepared as an account of work sponsored by an agency of the United States Government. Neither the United States Government nor any agency thereof, nor any of their employees, makes any warranty, express or implied, or assumes any legal liability or responsibility for the accuracy, completeness, or usefulness of any information, apparatus, product, or process disclosed, or represents that its use would not infringe privately owned rights. Reference herein to any specific commercial product, process, or service by trade name, trademark, manufacturer, or otherwise does not necessarily constitute or imply its endorsement, recommendation, or favoring by the United States Government or any agency thereof.

\section{CRediT authorship contribution statement}
\noindent GP: Conceptualization, Methodology, Investigation, Software, Writing - original draft. UV: Conceptualization, Methodology, Writing - review \& editing. DAH: Conceptualization, Data curation, Writing - review \& editing. TEY: Conceptualization, Writing - review \& editing. KW: Conceptualization, Supervision, Writing - review \& editing.

%% file: sections/appendix.tex
\appendix

\newcommand{\ltwo}{L^2(\Omega)}
\newcommand{\hone}{H^1(\Omega)}
\newcommand{\ltwoltwo}{L^2(L^2(\Omega); (t_0, t_f))}
\newcommand{\ltwohone}{L^2(H^1(\Omega); (t_0, t_f))}
\newcommand{\ltwohonenot}{L^2(H^1_0(\Omega); (t_0, t_f))}
\newcommand{\ltwohminusone}{L^2(H^{-1}(\Omega); (t_0, t_f))}

\newcommand{\intomegat}{\int_{t_0}^{t_f}\int_\Omega}
\newcommand{\dx}{\,dx}
\newcommand{\ds}{\,ds}
\newcommand{\dt}{\,dt}

\newcommand{\lagg}{\mathcal{L}^g}
\newcommand{\lagh}{\mathcal{L}^H}

\newcommand{\mhat}{\widehat{m}}
\newcommand{\mtilde}{\widetilde{m}}
\newcommand{\phat}{\widehat{p}}
\newcommand{\ptilde}{\widetilde{p}}
\newcommand{\uhat}{\widehat{u}}
\newcommand{\utilde}{\widetilde{u}}

\section{Model derivatives}
\label{sec:appendix-derivatives}

We develop the variational formulation for the reaction-diffusion model with therapy presented in Section~\ref{sec:models}. We will leverage the Lagrangian formalism to compute the adjoint and gradient expression for the model equations \cite{troltzsch2010optimal,manzoni2021optimal}. We will also compute the Hessian action for the model equations. Consider the governing equation \eqref{eqn:rd-model}, restated here for convenience,
\begin{equation*}
    \begin{alignedat}{2}
        \frac{\partial u}{\partial t} - \nabla\cdot\left(e^{m_D}\nabla u\right) - e^{m_\kappa} u(1-u) & = f(u) \quad &  & \text{in $\Omega\times (t_0, t_f)$}            \\
        u(x, t_0)                                                                      & = u_0     &  & \text{in $\Omega$}                       \\
        \nabla u \cdot \eta                                                                  & = 0       &  & \text{on $\partial\Omega\times (t_0, t_f)$}.
    \end{alignedat}
\end{equation*}
We will first fix notation. Here, \\

\begin{supertabular}{@{}l @{\quad\quad} l@{}}
    $(t_0, t_f)$ & The observation time window. \\
    $\Omega$ & The spatial domain in $\mathbb{R}^2$ or $\mathbb{R}^3$ with boundary $\partial\Omega$. \\
    $u(x,t)\in\mathcal{X}$ & The state, tumor volume fraction. \\
    $m_D(x)\in\hone$ & The diffusion parameter. \\
    $m_\kappa(x)\in\hone$ & The proliferation rate field. \\
    $f(x,u,t)\in\ltwohminusone$ & The source term. \\
    $v\in\mathcal{V}:=\ltwohone$ & A test function. \\
\end{supertabular} \\

The space for the state variable $\mathcal{X}:=\left\{ u \in \ltwohone\vert u_t \in \ltwohone\right\}$ is chosen to accommodate the required regularity for the spatial and temporal derivatives as well as the nonlinear term. Here the temporal derivative is denoted $u_t := \partial u / \partial t$. For control of the quadratic nonlinearity, we require $u\in L^2(L^4(\Omega) \cap H^1(\Omega); (0, t_f))$ to ensure that $u^2\in \ltwoltwo$. However, appealing to a Sobolev Embedding Theorem, the embedding $H^1(\Omega)\hookrightarrow L^4(\Omega)$ is continuous so that $u\in\ltwohone$ is sufficient (see Sec. 4.2.1. and Theorem 7.1 in \cite{troltzsch2010optimal}).

\subsection{Variational Formulation}
The weak form of the reaction-diffusion equation is obtained by multiplying the governing equation \eqref{eqn:rd-model} by a test function $v$ and integrating over the space-time domain $\Omega\times(0,t_f)$.
\begin{equation*}
    \intomegat u_t v\dx\dt - \intomegat \nabla\cdot(e^{m_D}\nabla u) v\dx\dt - \intomegat e^{m_\kappa} u(1-u)\dx\dt - \intomegat f v\dx\dt = 0
\end{equation*}
Straightforward application of the Divergence theorem along with the homogeneous Neumann boundary conditions yields the weak form,
\begin{equation*}
    \begin{aligned}
         \intomegat u_t v\dx\dt + \intomegat e^{m_D}\nabla u\cdot\nabla v\dx\dt & - \int_{t_0}^{t_f} \int_{\partial\Omega} \eta\cdot(e^{m_D}\nabla u)v \ds\dt \\
         & - \intomegat e^{m_\kappa} u(1-u) v - f v \dx\dt  = 0 \\
    \end{aligned}
\end{equation*}

\begin{equation*}
    \intomegat u_t v \dx\dt + \intomegat e^{m_D} \nabla u \cdot  \nabla v \dx\dt - \intomegat e^{m_\kappa} u(1-u) v \dx\dt - \intomegat f v \dx\dt = 0.
\end{equation*}

The development proceeded with a generic source term to account for therapy, but with the treatment model \eqref{eqn:treatment-model} given by \eqref{eqn:radio} and \eqref{eqn:chemo}, one would have the last term given by
\begin{equation*}
    \intomegat fv\dx\dt = \intomegat \kappa (1-S_{\text{rt}}(z))\,u\,v\dx\dt + \intomegat \alpha_{\text{ct}} \sum_{k} \exp(-\beta_{\text{ct}} (t-\tau_{k,\text{ct}})) u\,v\dx\dt
\end{equation*}
with appropriate application in time according to the therapy schedules $\mathcal{T}_{\text{rt}}$ and $\mathcal{T}_{\text{ct}}$.

\subsection{Lagrangian}
It is assumed that the state variable comes from a space with sufficient regularity, i.e.,  $\mathcal{X}$ or $W^{1,1}_{2}(\Omega)$, and the parameters $m$ similarly come from a space with sufficient regularity $\mathcal{M}$, i.e., $\hone\times\hone$. The forward model is given by the \gls*{pde} and is a mapping $\mathcal{F}:\mathcal{X}\times\mathcal{M}\rightarrow\mathcal{Y}$. The adjoint variable $p\in \mathcal{P} := \mathcal{Y}^*$, the dual of the action of the forward model. A suitable space is the previously defined $\mathcal{V}$ or $W^{1,1}_2(\Omega)$. For an extended discussion, the reader is referred to Chapter 3 of \cite{troltzsch2010optimal}. The inverse problem for the computation of the \gls*{map} point \eqref{eqn:map} is stated as:
\begin{equation*}
    \min_{m\in\mathcal{M}} \phi(m_D,m_\kappa) \quad , \quad \phi(m_D,m_\kappa) := -\log \nu_{\text{post}} (m \vert \boldsymbol{d}) = \underbrace{\frac{1}{2}\intomegat (\mathcal{B}u - \boldsymbol{d})^2 \dx\dt}_{\text{data misfit}} + \underbrace{\mathcal{R}(m)}_{\text{regularization}}.
\end{equation*}
While the data misfit functional is specified, the exact form of the regularization functional is not, as we prefer to derive the action in generality. Moreover, the observation operator $\mathcal{B}$ may incorporate a collection of Dirac deltas in time, to compare the state with the finitely many data collected over the simulation window. For the Gaussian random field priors developed in Section~\ref{sec:bip} analysis is available in the literature, in particular \cite{bui2013computational,villa2021hippylib}.

\subsection{Gradient expression}
Here we derive expressions for the gradient of $\phi$ with respect to the model parameters $m_D$ and $m_\kappa$ using the formal Lagrange method. We first form the Lagrangian functional, $\lagg$ (where the superscript $g$ denotes the role in deriving the gradient), that combines the regularized data misfit $\phi(m)$ with the weak form of the model equations. The Lagrangian functional is,
\begin{equation*}
    \begin{split}
        \lagg(u, p, m_D, m_\kappa) &:= \frac{1}{2}\intomegat (\mathcal{B}u - d)^2 \dx\dt + \alpha \mathcal{R}(m_D, m_\kappa) \\
        &+ \intomegat \bigg[ u_t p + e^{m_D} \nabla u \cdot \nabla p - e^{m_\kappa} u(1-u) p - f(u) p \bigg] \dx\dt
    \end{split}
\end{equation*}
for functions $(u,p,(m_D,m_\kappa))\in\mathcal{X}\times\mathcal{P}\times\mathcal{M}$.

\subsubsection{The variational formulation}
Taking variations of $\lagg$ with respect to $p\in\mathcal{P}$ and requiring them to vanish for all admissible variations $\phat$ simply recovers the weak form of the forward reaction-diffusion model previously derived. That is, $\delta_p \lagg = 0$ for all $\phat\in\mathcal{P}$ yields the weak form of the \gls*{pde}.
\begin{equation*}
    \intomegat \frac{\partial u}{\partial t}\phat - e^{m_D} \nabla u\cdot\nabla\phat - e^{m_\kappa} u(1-u)\phat - f(u) \phat \dx\dt = 0
\end{equation*}

\subsubsection{The adjoint equation}
Next, we require that variations of $\lagg$ with respect to the state $u$ vanish for all admissible variations $\uhat\in \mathcal{X}$, that is $\delta_u \lagg = 0$ for all $\uhat\in\mathcal{X}$. This will yield the weak form of the adjoint \gls*{pde}. We have,
\begin{equation*}
    \intomegat (\mathcal{B}u - d)(\mathcal{B}\uhat)\dx\dt + \intomegat \bigg[ p\frac{\partial\uhat}{\partial t} + e^{m_D}\nabla \uhat \cdot\nabla p - e^{m_\kappa} p\uhat + 2 e^{m_\kappa} p u\uhat \bigg] \dx\dt - \intomegat \delta_u f(u) \uhat\,p\,\dx\dt
\end{equation*}
for all $\uhat\in\mathcal{X}$. For the specific treatment models that we consider, we have
\begin{equation*}
    \intomegat \delta_u f(u) \,\uhat\,p\,\dx\dt = \intomegat \kappa (1-S_{\text{rt}}(z))\,\uhat\,p\dx\dt + \intomegat \alpha_{\text{ct}} \sum_{k} \exp(-\beta_{\text{ct}} (t-\tau_{k,\text{ct}})) \uhat\,p\dx\dt
\end{equation*}

By appropriate integration by parts in time and space to remove derivatives of $\uhat$ and arguing the arbitrariness of $\uhat$, we can arrive at the strong form of the adjoint equation, a terminal boundary value problem given by,
\begin{equation*}
    \begin{alignedat}{2}
        -\frac{\partial p}{\partial t} - &\nabla\cdot\left( e^{m_D} \nabla p \right) - \left(e^{m_\kappa} + 2e^{m_\kappa} u\right) p & & \\
        & = -\mathcal{B}^*(\mathcal{B}u - d) - \kappa (1-S_{\text{rt}}(z)) p - \alpha_{\text{ct}} \sum_{k} \exp(-\beta_{\text{ct}} (t-\tau_{k,\text{ct}})) p \quad &  & \text{in $\Omega\times (t_0, t_f)$} \\
        p(x, t_f) & = 0 & & \text{in $\Omega$} \\
        e^{m_D}\nabla p \cdot \eta & = 0 & & \text{on $\partial\Omega\times (t_0, t_f)$}
    \end{alignedat}
\end{equation*}

\subsubsection{The gradient expression}
Finally, we derive expressions for the gradient, the Fr\'echet derivative of $\phi$ with respect to $m_D$ and $m_\kappa$, denoted $\mathfrak{D}_{m_D}\phi$ and $\mathfrak{D}_{m_\kappa}\phi$, respectively. We consider the variations of the Lagrangian with respect to the parameters. The Fr\'echet derivative of $\phi(m_D, m_\kappa)$ with respect to $m_D$ in an arbitrary direction $\widehat{m}\in\hone$ evaluated at $(m_D, m_\kappa)$ is given by $\delta_{m_D}\lagg$, that is
\begin{equation*}
    \mathfrak{D}_{m_D} \phi(m_D, m_\kappa, \mhat_d) :=  \delta_{m_D} \mathcal{R}(m_D, m_\kappa) + \intomegat \widehat{m}_D e^{m_D} \nabla u \cdot \nabla p \dx\dt.
\end{equation*}
Similarly, the Fr\'echet derivative of $\phi(m_D, m_\kappa)$ with respect to $m_\kappa$ in an arbitrary direction $\widehat{m}_2\in\hone$ evaluated at $(m_D, m_\kappa)$ is given by $\delta_{m_\kappa}\lagg$,
\begin{equation*}
    \mathfrak{D}_{m_\kappa} \phi(m_D, m_\kappa, \mhat_\kappa) := \delta_{m_\kappa} \alpha\mathcal{R}(m_D, m_\kappa) + \intomegat \mhat_\kappa e^{m_\kappa} u(1-u)p \dx\dt.
\end{equation*}
The gradient $\mathfrak{G}$ with respect to a parameter $m$ is defined as the Riesz representer of the Fr\'echet derivative of $\phi$ with respect to a chosen inner product,
\begin{equation*}
    (\mathfrak{G}_m(m), \mhat) := \mathfrak{D}_m \phi(m, \mhat) = \delta_m \lagg.
\end{equation*}
In summary, to compute the gradient at $(m_D, m_\kappa)$, we
\begin{enumerate}
    \item Solve the forward model equation for $u$, given $(m_D, m_\kappa)$.
    \item Solve the adjoint model equation for $p$, given $(m_D, m_\kappa)$ and $u$.
    \item Evaluate the Fr\'echet derivatives $\delta_{m_D}\lagg$ and $\delta_{m_\kappa}\lagg$, given $(m_D, m_\kappa)$, $u$, and $p$.
\end{enumerate}

\subsection{Hessian action}
We now form the Lagrangian for the Hessian, $\lagh$ (where the superscript $H$ refers to the role in deriving the Hessian action). We retain the notation that $\tilde{\cdot}$ denotes a trial function and $\hat{\cdot}$ denotes a test function. We consider the Hessian action in an arbitrary direction $(\mtilde_D, \mtilde_\kappa)$, replacing $(\mhat_D, \mhat_\kappa)$ in the previous section. This is done to preserve the notation $\widehat{\cdot}$ for the \emph{current} variations. Similarly, we replace $\phat$ with $\ptilde$ and $\uhat$ with $\utilde$. We refer to $\utilde$ and $\ptilde$ as the \emph{incremental state} and \emph{incremental adjoint} variables, respectively. The Lagrangian functional for the Hessian is,
\begin{equation*}
    \begin{split}
        \lagh(u,p,&m_D,m_\kappa,\utilde,\ptilde,\mtilde_D,\mtilde_\kappa) := \\
        &\underbrace{\delta_{m_D}R(m_D, m_\kappa) + \intomegat \mtilde_D e^{m_D}\nabla u\cdot \nabla p \dx\dt}_{\text{Fr\'echet derivative with respect to } m_D \text{ in direction } \mtilde_D}   \\
        &+ \underbrace{\delta_{m_\kappa}R(m_D, m_\kappa) + \intomegat \mtilde_\kappa e^{m_\kappa} u(1-u)p \dx\dt}_{\text{Fr\'echet derivative with respect to } m_\kappa \text{ in direction } \mtilde_\kappa} \\
        &+ \underbrace{\intomegat \bigg[ \frac{\partial u}{\partial t}\ptilde + e^{m_D}\nabla u\cdot \nabla \ptilde - e^{m_\kappa} u(1-u)\ptilde - f(u)\ptilde \bigg] \dx\dt}_{\text{Weak form of forward model equation}} \\
        &+ \underbrace{\intomegat \bigg[ (\mathcal{B} u - d) \mathcal{B} \utilde + p\frac{\partial \utilde}{\partial t} + e^{m_D}\nabla \utilde \cdot \nabla p - e^{m_\kappa} \utilde p + 2m_\kappa u\utilde p \bigg] \dx\dt + \delta_u f(u)}_{\text{Weak form of adjoint model equation}},
    \end{split}
\end{equation*}
where $(u,p,(m_D,m_\kappa),\utilde,\ptilde,(\mtilde_D,\mtilde_\kappa))\in\mathcal{X}\times\mathcal{P}\times\mathcal{M}\times\mathcal{X}\times\mathcal{P}\times\mathcal{M}$. To derive the expression for the action of the Hessian of $\phi$ with respect to $(m_D,m_\kappa)$ in a direction $(\mtilde_D,\mtilde_\kappa)$ we take variations of $\lagh$ with respect to its arguments. \\

\subsubsection{The incremental forward equation}
Requiring variations of $\lagh$ with respect to the adjoint $p$ to vanish for all admissible variations $\phat\in\mathcal{P}$ yields the \emph{incremental forward} equation: Given $(m_D,m_\kappa)$, $(\mtilde_D,\mtilde_\kappa)$, and $u$, find the incremental state $\utilde$ such that for all $\phat\in\mathcal{P}$,
\begin{equation*}
    \intomegat \bigg[ \mtilde_D e^{m_D}\nabla u \cdot \nabla \phat + \mtilde_\kappa e^{m_\kappa} u(1-u)\phat - e^{m_\kappa} \utilde\,\phat + 2e^{m_\kappa} u\,\utilde\,\phat + \phat \frac{\partial \utilde}{\partial t} + e^{m_D} \nabla \utilde \cdot \nabla \phat \bigg] \dx\dt = 0.
\end{equation*}
Integration by parts to remove derivatives of $\phat$ and arguing the arbitrariness of $\phat$ yields the strong form of the incremental forward equation,
\begin{equation*}
    \begin{alignedat}{2}
        \frac{\partial \utilde}{\partial t} - \nabla\cdot\left( e^{m_D} \nabla \utilde \right) - e^{m_\kappa} \utilde + 2e^{m_\kappa} u\,\utilde & = \mtilde_D \nabla \cdot (e^{m_D}\nabla u) - \mtilde_\kappa e^{m_\kappa}\, u (1-u) \quad &  & \text{in $\Omega\times (t_0, t_f)$}            \\
        \utilde(x, t_0) & = 0     &  & \text{in $\Omega$}                       \\
        e^{m_D}\nabla \utilde \cdot \eta & = 0 & & \text{on $\partial\Omega\times (t_0, t_f)$}
    \end{alignedat}
\end{equation*}
Note that the initial condition is identically zero for the incremental state variable in order to satisfy first order optimality conditions (see Theorem 2.22 \cite{troltzsch2010optimal}). \\

\subsubsection{The incremental adjoint equation}
Taking variations of $\lagh$ with respect to the state $u$ and requiring them to vanish for all admissible variations yields the incremental adjoint problem, that is $\delta_u \lagh = 0$ for all $\uhat\in\mathcal{X}$,
\begin{equation*}
    \begin{split}
        &\intomegat \mathcal{B}\uhat\,\mathcal{B}\utilde + 2e^{m_\kappa}\uhat\,\utilde \dx \dt \\
        &+ \intomegat \bigg[ \frac{\partial \uhat}{\partial t}\ptilde + e^{m_D} \nabla \uhat \cdot \nabla \phat - e^{m_\kappa} \uhat\,(1-u)\,\ptilde + e^{m_\kappa} u\,\uhat\,\ptilde \bigg] \dx\dt \\
        &+ \intomegat \bigg[ \mtilde_\kappa e^{m_\kappa} \uhat\,(1-u)\,p - \mtilde_\kappa e^{m_\kappa} \,u \uhat\, p + \mtilde_D e^{m_D}\nabla \uhat \cdot \nabla p \bigg] \dx\dt = 0
    \end{split}
\end{equation*}
Note that the source terms considered in this work are linear with respect to the state variable and so $\delta_{uu} f(u) = 0$. However, for treatment terms with nonlinear state dependence, one should take care to account for the required term. Appropriate integration by parts in time and space to remove derivatives of $\uhat$ and arguing the arbitrariness of $\uhat$ yields the strong form of the incremental adjoint equation,
\begin{equation*}
    \begin{alignedat}{2}
        -\frac{\partial \ptilde}{\partial t} -  &\mtilde_D \nabla\cdot\left( e^{m_D} \nabla \ptilde \right) + e^{m_\kappa} (2u-1)\ptilde  \\
        & = -\mathcal{B}^*(\mathcal{B}\utilde) - 2e^{m_\kappa} \utilde + \mtilde_D\nabla \cdot (e^{m_D}\nabla p) - \mtilde_\kappa e^{m_\kappa} p(1-2u) \quad &  & \text{in $\Omega\times (t_0, t_f)$} \\
        \ptilde(x, t_f) & = 0 & & \text{in $\Omega$} \\
        e^{m_D}\nabla \ptilde \cdot \eta & = 0 &  & \text{on $\partial\Omega\times (t_0, t_f)$}
    \end{alignedat}
\end{equation*}

\subsubsection{The Hessian action}
Finally, we derive expressions for the action of the Hessian of $\phi$ with respect to $(m_0, m_\kappa)$ in a direction $(\mtilde_0, \mtilde_\kappa)$, that is, the second Fr\'echet derivative of $\phi$, $\mathfrak{D}^2 \phi$. Since we have multiple parameter fields, the Hessian is a block operator and we must consider the mixed derivatives,
\begin{equation*}
    \mathcal{H}(m_D, m_\kappa) := \begin{bmatrix} \mathcal{H}_{DD} & \mathcal{H}_{D\kappa} \\ \mathcal{H}_{\kappa D} & \mathcal{H}_{\kappa\kappa} \end{bmatrix}.
\end{equation*}
Given $(m_D, m_\kappa)$, the action of the first block row of $\mathcal{H}$ in the direction $(\mtilde_D, \mtilde_\kappa)$ is given by $\delta_{m_D}\lagh$,
\begin{equation*}
    \begin{split}
        \delta_{m_D}\lagh &:= (\mhat_d, \mathcal{H}_{dd}\mtilde_D) + (\mhat_\kappa, \mathcal{H}_{d\kappa}\mtilde_\kappa) \\
        &= \alpha \delta^2_{m_D} \mathcal{R}(m_D, m_\kappa) + \alpha \delta_{m_D} (\delta_{m_\kappa} \mathcal{R}(m_D, m_\kappa)) \\
        &+ \intomegat \bigg[ \mtilde_D (\mhat_de^{m_D}\nabla u \cdot \nabla p) + \mhat_d e^{m_D}\nabla \utilde \cdot \nabla p + \mhat_d e^{m_D} \nabla u \cdot \nabla \ptilde \bigg] \dx\dt
    \end{split}
\end{equation*}
Similarly, given $(m_D, m_\kappa)$, the action of the second block row of $\mathcal{H}$ in the direction $(\mtilde_D, \mtilde_\kappa)$ is given by $\delta_{m_\kappa}\lagh$,
\begin{equation*}
    \begin{split}
        \delta_{m_\kappa}\lagh &:= (\mhat_\kappa, \mathcal{H}_{\kappa d}\mtilde_D) + (\mhat_\kappa, \mathcal{H}_{\kappa\kappa}\mtilde_\kappa) = \delta_{m_\kappa}\lagh\\
        &= \delta_{m_\kappa}(\delta_{m_D} \mathcal{R}(m_D, m_\kappa)) +  \intomegat \bigg[ \mhat_\kappa e^{m_\kappa} (2\utilde\,u - 1) - \mhat_\kappa e^{m_\kappa} u\,(1-u)\,\ptilde \bigg] \dx\dt
    \end{split}
\end{equation*}
In summary, to compute the Hessian action at a point in parameter space $(m_D, m_\kappa)\in\mathcal{M}$ in a direction $(\mtilde_D, \mtilde_\kappa)$, we
\begin{enumerate}
    \item Solve the forward model for $u$, given $(m_D, m_\kappa)$.
    \item Solve the adjoint equation for $p$, given $(m_D, m_\kappa)$ and $u$.
    \item Solve the incremental forward equation for $\utilde$, given $(m_D, m_\kappa)$, $(\mtilde_D, \mtilde_\kappa)$, and $u$.
    \item Solve the incremental adjoint equation for $\ptilde$, given $(m_D, m_\kappa)$, $(\mtilde_D, \mtilde_\kappa)$, $u$, $p$, and $\utilde$.
    \item Evaluate the Hessian actions, given $(m_D, m_\kappa)$, $(\mtilde_D, \mtilde_\kappa)$, $u$, $p$, $\utilde$, and $\ptilde$.
\end{enumerate}

\section{Additional scaling studies}
\label{sec:appendix-scaling}
Since the computational domain developed in Sec.~\ref{sec:data} is assumed to be fixed at simulation time, we are primarily concerned with the strong scalability of the forward solve. That is, how quickly can a given amount of work be parallelized and solved. While it is not currently useful to refine the mesh past the resolution of the data, higher quality data may be available, thus requiring meshes with additional \glspl*{dof} to accurately represent the data. To account for our implementation's ability to handle these cases we also perform a weak scaling study.

\subsection{Strong scaling of the adjoint problem}
\begin{figure}[!htb]
    \centering
    \subfloat[Intra-node]{\includegraphics[width=0.49\linewidth]{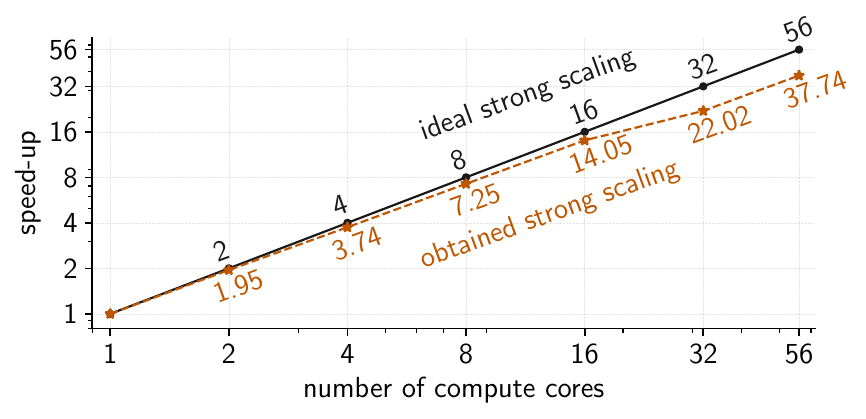}}
    \subfloat[Inter-node]{\includegraphics[width=0.49\linewidth]{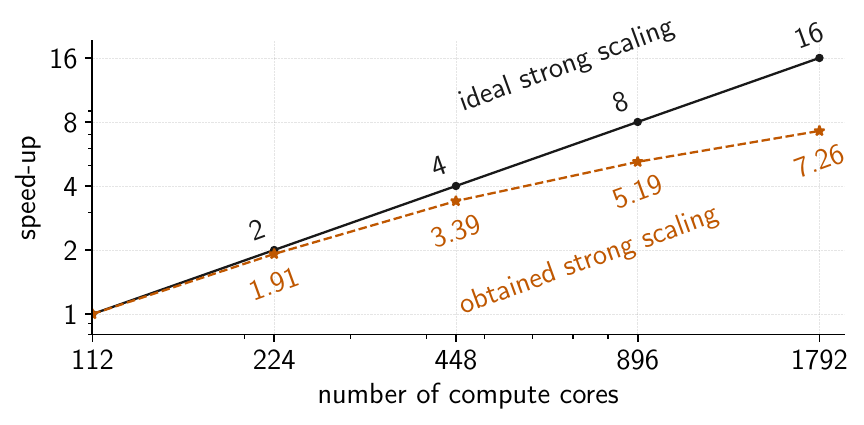}}
    \caption{Strong scaling of the adjoint solve in (a) 14.2 seconds on one node for a benchmark problem with 1 million \glspl*{dof} and (b) 4.1 seconds on 32 nodes for a benchmark problem with 4 million \glspl*{dof}.}
    \label{fig:adjoint-strong-scaling}%
\end{figure}

In Fig.~\ref{fig:adjoint-strong-scaling}, we report the strong scaling of the adjoint solve for the benchmark problem outlined in Sec.~\ref{sec:implementation}. Note that the inter-node study requires at least two nodes because of the required memory footprint for the adjoint solve (storage of the state variable). Once more we observe excellent scaling within node and adequate scaling out of node.

\subsection{Weak scaling}
\begin{figure}[!htb]
    \centering
    \includegraphics[width=0.8\linewidth]{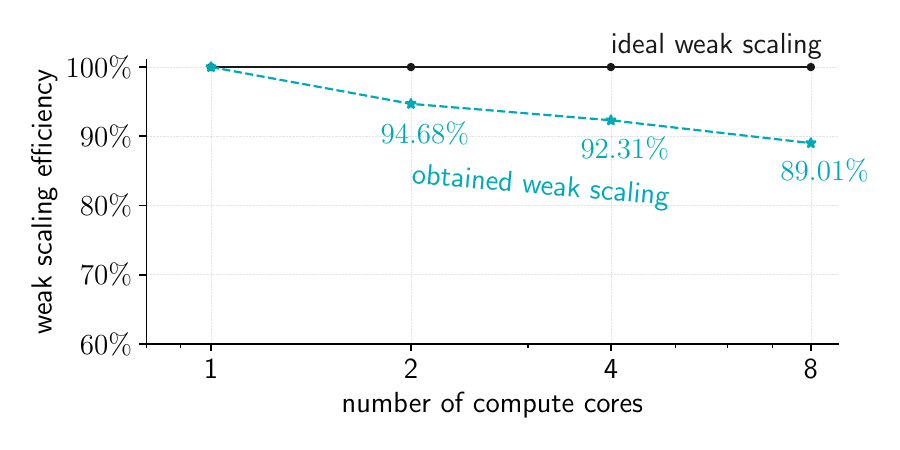}
    \caption{Weak scaling of the forward solve with 300,000 \glspl*{dof} per processor on a Frontera CLX node.}
    \label{fig:weak-scaling}
\end{figure}

For the weak scaling study, we fix the amount of work per processor to be 300,000 \glspl*{dof} and test the efficiency by adding additional work and cores in equal proportion until resources are exhausted. Results on a single Frontera CLX node are reported in Fig.~\ref{fig:weak-scaling}, where we observe a reasonably efficient scaling up to 2.4 million \glspl*{dof}.

\section{Model validation summary statistics}
\label{sec:appendix-summary-statistics}
We collect and tabulate summary statistics for the clinical model validation study presented in Section~\ref{sec:ivygap_performance}. In particular, we report the mean, standard deviation, and 90\% credible interval (5$^{th}$ to 95$^{th}$ percentile) for both prior distribution and posterior distribution pushforwards.

\begin{table}[!htb]
\centering
\caption{Summary of mean, standard deviation, and 90\% credible interval by patient for the Dice similarity coefficient in the last-to-final prediction case.}
\begin{tabular}{|c|ccc|ccc|}
\hline
\multirow{2}{*}{\textbf{Patient}} & \multicolumn{3}{c|}{\textbf{Prior Distribution}} & \multicolumn{3}{c|}{\textbf{Laplace Approximation}} \\ \cline{2-7} 
 & \textbf{Mean} & \textbf{Std. Dev.} & \textbf{Credible Interval} & \textbf{Mean} & \textbf{Std. Dev.} & \textbf{Credible Interval} \\ \hline
W03 & 0.566 & 0.018 & (0.538, 0.593) & 0.462 & 0.0022 & (0.459, 0.466) \\ \hline
W11 & 0.447 & 0.028 & (0.401, 0.490) & 0.731 & 0.0027 & (0.727, 0.736) \\ \hline
W16 & 0.879 & 0.0046 & (0.872, 0.886) & 0.870 & 0.0018 & (0.867, 0.873) \\ \hline
W29 & 0.417 & 0.016 & (0.390, 0.443) & 0.467 & 0.0062 & (0.457, 0.476) \\ \hline
W35 & 0.427 & 0.032 & (0.369, 0.481) & 0.636 & 0.0052 & (0.627, 0.644) \\ \hline
W36 & 0.671 & 0.030 & (0.619, 0.715) & 0.654 & 0.0031 & (0.649, 0.659) \\ \hline
W43 & 0.594 & 0.010 & (0.577, 0.609) & 0.597 & 0.0018 & (0.595, 0.600) \\ \hline
W53 & 0.095 & 0.020 & (0.070, 0.131) & 0.267 & 0.014 & (0.245, 0.288) \\ \hline
\end{tabular}
\end{table}

\begin{table}[!htb]
\centering
\caption{Summary of mean, standard deviation, and 90\% credible interval by patient for the Dice similarity coefficient in the initial-to-final prediction case.}
\begin{tabular}{|c|ccc|ccc|}
\hline
\multirow{2}{*}{\textbf{Patient}} & \multicolumn{3}{c|}{\textbf{Prior Distribution}} & \multicolumn{3}{c|}{\textbf{Laplace Approximation}} \\ \cline{2-7} 
 & \textbf{Mean} & \textbf{Std. Dev.} & \textbf{Credible Interval} & \textbf{Mean} & \textbf{Std. Dev.} & \textbf{Credible Interval} \\ \hline
W03 & 0.506 & 0.061 & (0.416, 0.616) & 0.438 & 0.031 & (0.387, 0.491) \\ \hline
W11 & 0.201 & 0.0015 & (0.201, 0.203) & 0.502 & 0.038 & (0.431, 0.551) \\ \hline
W16 & 0.019 & 0.060 & (0.000, 0.128) & 0.779 & 0.010 & (0.762, 0.794) \\ \hline
W29 & 0.272 & 0.060 & (0.192, 0.362) & 0.285 & 0.023 & (0.252, 0.327) \\ \hline
W35 & 0.178 & 0.046 & (0.137, 0.268) & 0.559 & 0.013 & (0.536, 0.572) \\ \hline
W36 & 0.398 & 0.088 & (0.269, 0.540) & 0.477 & 0.063 & (0.374, 0.576) \\ \hline
W43 & 0.285 & 0.186 & (0.004, 0.546) & 0.501 & 0.0053 & (0.493, 0.509) \\ \hline
W53 & 0.067 & 0.0024 & (0.066, 0.070) & 0.146 & 0.021 & (0.112, 0.180) \\ \hline
\end{tabular}
\end{table}

\begin{table}[!htb]
\centering
\caption{Summary of mean, standard deviation, and 90\% credible interval by patient for the relative error in total tumor cellularity in the last-to-final prediction case.}
\begin{tabular}{|c|ccc|ccc|}
\hline
\multirow{2}{*}{\textbf{Patient}} & \multicolumn{3}{c|}{\textbf{Prior Distribution}} & \multicolumn{3}{c|}{\textbf{Laplace Approximation}} \\ \cline{2-7} 
 & \textbf{Mean} & \textbf{Std. Dev.} & \textbf{Credible Interval} & \textbf{Mean} & \textbf{Std. Dev.} & \textbf{Credible Interval} \\ \hline
W03 & 0.122 & 0.151 & (-0.085, 0.391) & -0.137 & 0.0064 & (-0.147, -0.125) \\ \hline
W11 & 5.041 & 0.715 & (3.949, 6.296) & 1.683 & 0.012 & (1.664, 1.704) \\ \hline
W16 & 0.046 & 0.053 & (-0.039, 0.133) & 0.683 & 0.0077 & (0.670, 0.696) \\ \hline
W29 & 1.640 & 0.383 & (1.088, 2.294) & 1.971 & 0.075 & (1.855, 2.102) \\ \hline
W35 & 5.891 & 0.781 & (4.749, 7.344) & 3.146 & 0.042 & (3.079, 3.215) \\ \hline
W36 & -0.222 & 0.239 & (-0.550, 0.222) & -0.162 & 0.0077 & (-0.175, -0.150) \\ \hline
W43 & 1.935 & 0.213 & (1.617, 2.301) & 2.766 & 0.021 & (2.731, 2.800) \\ \hline
W53 & 46.081 & 12.865 & (26.609, 68.141) & 11.059 & 0.567 & (10.226, 11.961) \\ \hline
\end{tabular}
\end{table}

\begin{table}[!htb]
\centering
\caption{Summary of mean, standard deviation, and 90\% credible interval by patient for the relative error in total tumor cellularity in the initial-to-final prediction case.}
\begin{tabular}{|c|ccc|ccc|}
\hline
\multirow{2}{*}{\textbf{Patient}} & \multicolumn{3}{c|}{\textbf{Prior Distribution}} & \multicolumn{3}{c|}{\textbf{Laplace Approximation}} \\ \cline{2-7} 
 & \textbf{Mean} & \textbf{Std. Dev.} & \textbf{Credible Interval} & \textbf{Mean} & \textbf{Std. Dev.} & \textbf{Credible Interval} \\ \hline
W03 & 2.863 & 1.496 & (0.555, 5.394) & 1.175 & 0.341 & (0.685, 1.774) \\ \hline
W11 & 20.925 & 0.764 & (19.197, 21.370) & 4.177 & 0.481 & (3.665, 5.182) \\ \hline
W16 & -0.994 & 0.029 & (-1.000, -0.973) & 1.637 & 0.049 & (1.562, 1.721) \\ \hline
W29 & 2.966 & 2.773 & (-0.525, 8.284) & 8.449 & 1.703 & (6.016, 11.436) \\ \hline
W35 & 19.379 & 6.115 & (9.380, 29.660) & 4.244 & 0.136 & (4.073, 4.425) \\ \hline
W36 & 1.270 & 1.198 & (-0.430, 3.541) & 0.966 & 0.420 & (0.464, 1.785) \\ \hline
W43 & -0.376 & 0.705 & (-0.996, 1.149) & 4.618 & 0.107 & (4.444, 4.797) \\ \hline
W53 & 85.526 & 5.741 & (73.815, 89.354) & 27.216 & 4.374 & (21.167, 35.063) \\ \hline
\end{tabular}
\end{table}